\documentclass[11pt]{article}

\usepackage[preprint]{acl}

\usepackage{times}
\usepackage{latexsym}

\usepackage[T1]{fontenc}

\usepackage[utf8]{inputenc}

\usepackage{microtype}

\usepackage{inconsolata}

\usepackage{graphicx}

\usepackage{color-edits}
\addauthor{vc}{blue}
\addauthor{wc}{red}
\addauthor{ns}{orange}

\title{How Do Developers Edit AI-Generated Code, and How Well Do LLMs Infer Their Edits? Insights from \datasetSize Developer Edits}

\author{
Jenny T. Liang\footnotemark[1] \quad
Mihika Bairathi\footnotemark[1]\\
\textbf{Wayne Chi} \quad
\textbf{Ameet Talwakar} \quad
\textbf{Nishant Subramani} \quad \textbf{Valerie Chen}\\
\texttt{\small \{jtliang, mbairath, waynechi, atalwalk, nishant2, vchen2\}@cs.cmu.edu}\\
}

\usepackage{natbib}
\usepackage{url}
\usepackage{hyperref}
\usepackage{etoolbox}

\usepackage{microtype}

\usepackage{amssymb,bm,mathtools,color}

\usepackage[normalem]{ulem}
\usepackage{enumitem}

\usepackage{booktabs}
\usepackage{multirow}
\usepackage{tabularx}
\usepackage{multicol}
\usepackage{multirow}
\usepackage{colortbl}
\usepackage{hhline}
\usepackage{stfloats}
\usepackage[most]{tcolorbox}
\usepackage{multirow}
\usepackage{graphicx}
\usepackage{subcaption}
\usepackage{wrapfig}
\usepackage{etoolbox}
\usepackage{enumitem}
\usepackage{graphicx}
\usepackage{enumitem}
\usepackage[table]{xcolor}
\usepackage{xspace}
\usepackage[noabbrev,capitalize]{cleveref}
\usepackage{amsmath}
\usepackage{roboto}
\usepackage{enumitem}
\usepackage{fontawesome}

\newcolumntype{Y}{>{\centering\arraybackslash}X}

\usepackage{xcolor}
\definecolor{red}{rgb}{0.74,0.08,0.10}
\definecolor{green}{rgb}{0.26,0.49,0.18}
\definecolor{blue}{rgb}{0.22,0.53,0.75}

\usepackage{colortbl}
\definecolor{Gray}{gray}{0.9}
\definecolor{LightCyan}{rgb}{0.75,1,1}

\makeatletter
\newcommand\notsotiny{\@setfontsize\notsotiny\@viiipt\@ixpt}
\makeatother
\definecolor{PAblue}{RGB}{0,122,204}%

\newcommand{\ie}{\emph{i.e.}, }
\newcommand{\eg}{\emph{e.g.}, }

\newcommand{\makename}[3][s]{%
  \expandafter\newcommand\csname #2\endcsname{#3\xspace}%
  \expandafter\newcommand\csname #2s\endcsname{#3#1\xspace}%
}

\newcommand{\dataset}{{\sc{Decode}}\xspace}
\newcommand{\bigDataset}{{\sc{Decode-Real}}\xspace}
\newcommand{\smallDataset}{{\sc{Decode-Synthetic}}\xspace}

\newcommand{\datasetSize}{53.6K\xspace}

\newcommand{\modelName}[1]{{\roboto{#1}}}

\newcommand{\ieCustom}{{\emph{i.e.,~}}}
\newcommand{\egCustom}{{\emph{e.g.,~}}}

\definecolor{quoteBorder}{HTML}{5C76AD}
\definecolor{quoteBg}{HTML}{F2F3F8}
\newtcolorbox{pullquote}{
  colback=quoteBg,
  colframe=white,
  boxrule=0pt,
  arc=3pt,
  left=8pt,
  right=8pt,
  top=6pt,
  bottom=6pt,
  fontupper=\itshape,
  enhanced,
  borderline west={3pt}{0pt}{quoteBorder},
}

\newcommand{\icon}[1]{%
  \includegraphics[height=0.8em]{#1}\xspace
}

\newcommand{\clockIcon}{\icon{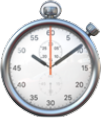}}

\newcommand{\robotIcon}{\icon{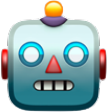}}

\newcommand{\metricDelta}[2]{{\color{#1} \tiny{(#2)}}}

\newcommand{\htmlicon}{\icon{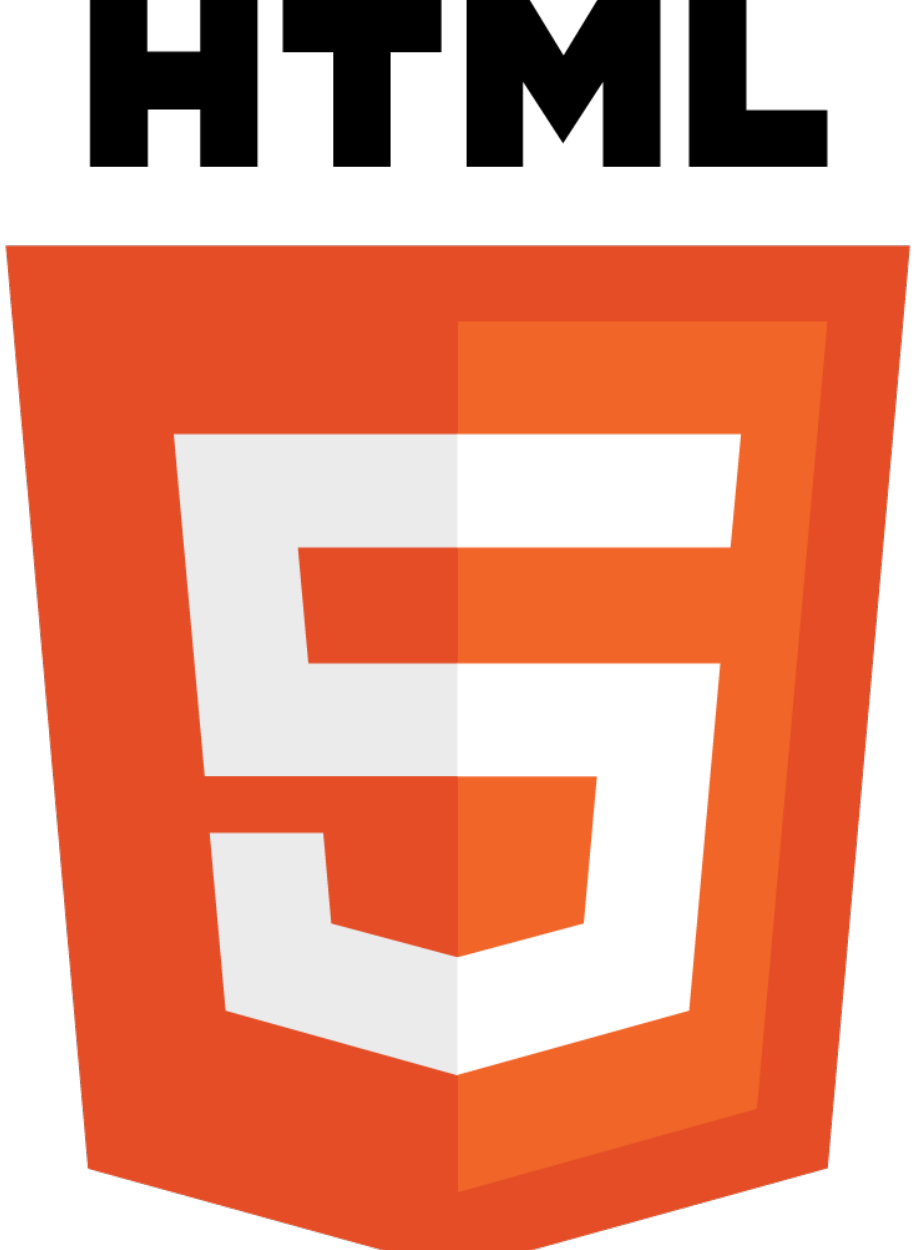}}
\newcommand{\javascripticon}{\icon{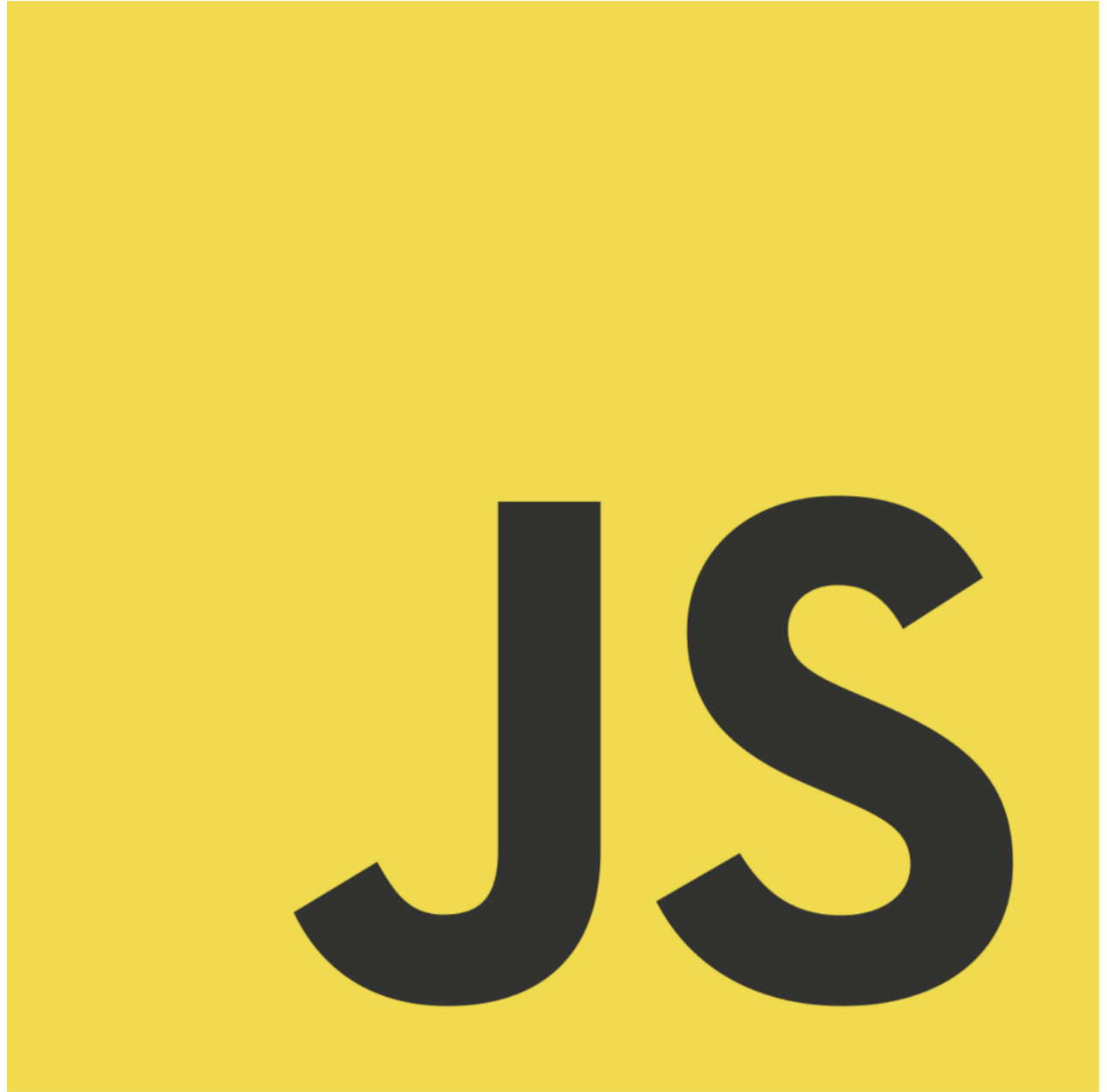}}
\newcommand{\typescripticon}{\icon{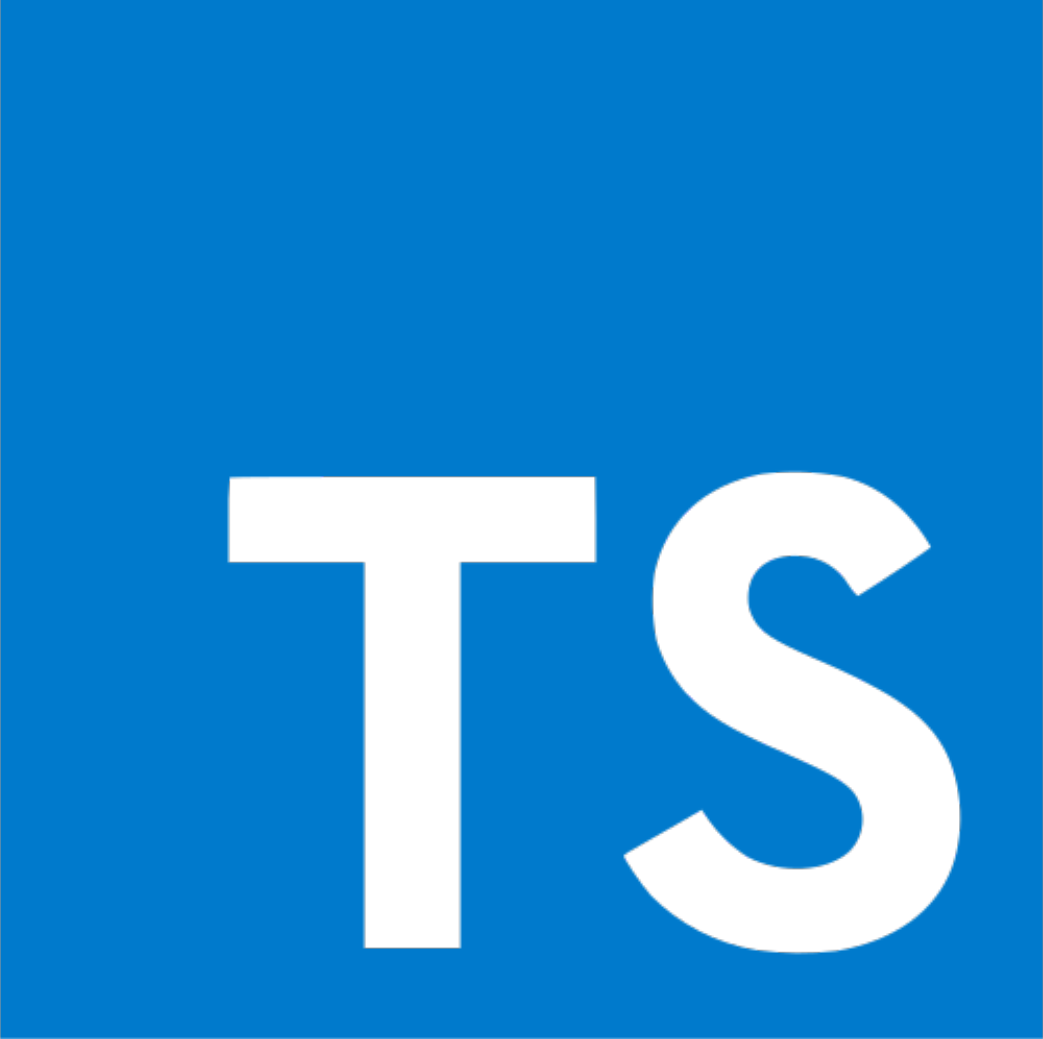}}
\newcommand{\pythonicon}{\icon{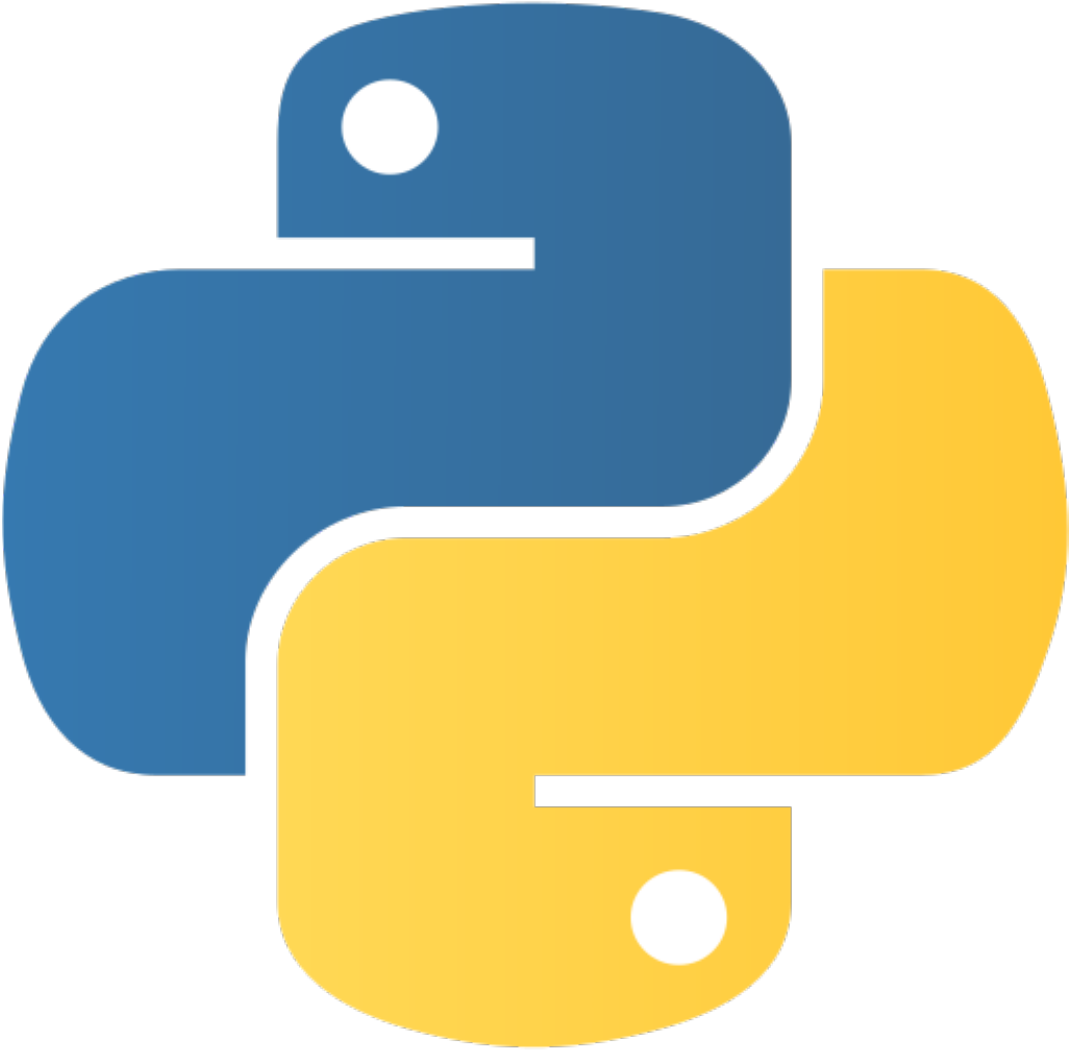}}
\newcommand{\coffeescripticon}{\icon{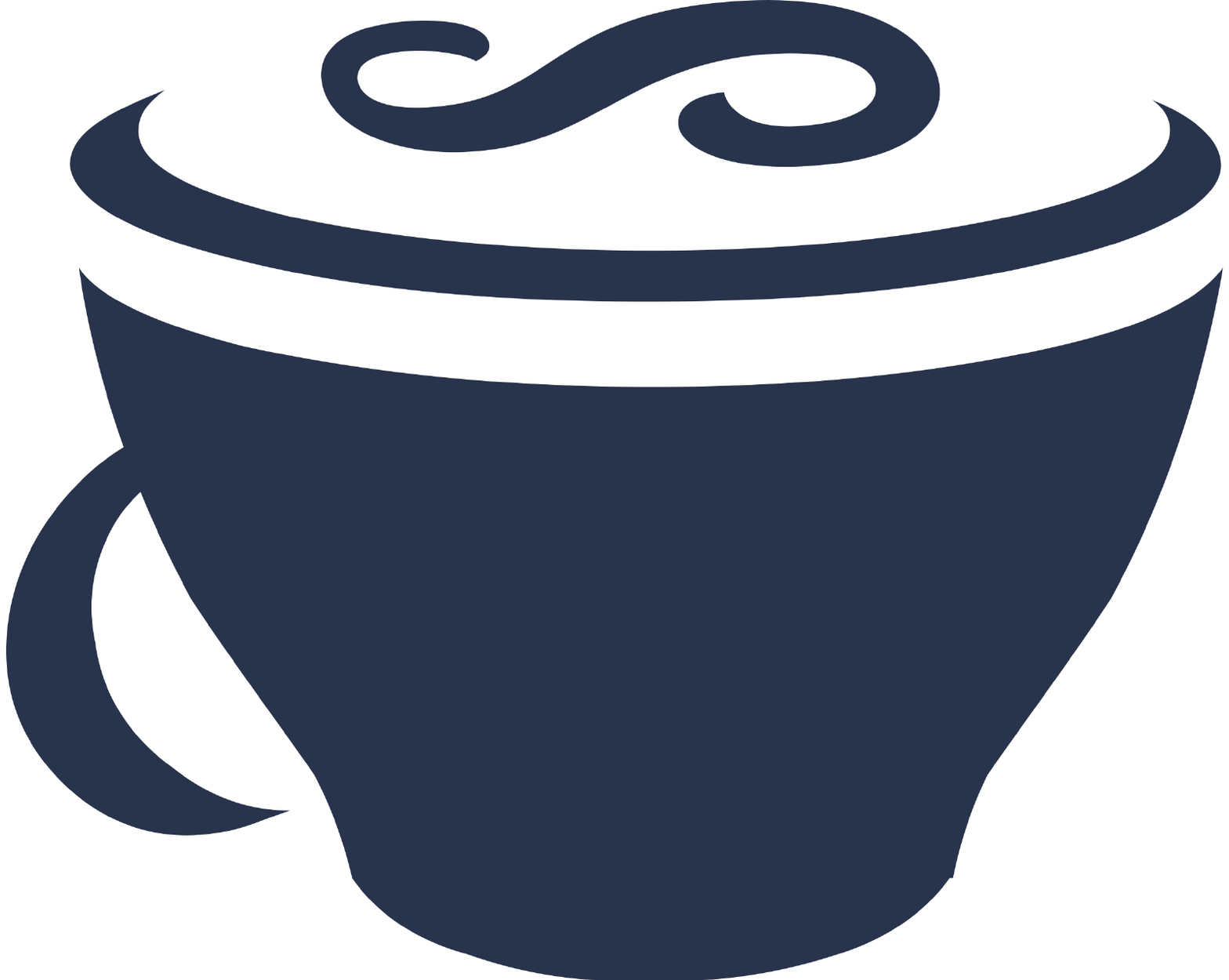}}

\definecolor{softgreen}{RGB}{85, 149, 90}
\definecolor{softred}{RGB}{193, 86, 86}

\newcommand{\cmark}{\textcolor{softgreen}{\faCheck}}
\newcommand{\xmark}{\textcolor{softred}{\faTimes}}

\begin{document}

\maketitle

\begin{abstract}
Imperfections in AI-generated code require that software developers modify generations through manual code edits.
However, little is known about how developers edit AI-generated code, and to what extent LLMs can infer implicit developer intent from this data, given the lack of high-quality, realistic code editing data.
This information is essential for designing and building usable AI-developer interactions.
To address this gap, we introduce \dataset (\textbf{\underline{D}}eveloper \textbf{\underline{E}}dits of \textbf{\underline{Co}}de \textbf{\underline{D}}atas\textbf{\underline{e}}t), a dataset of \datasetSize real-world in-IDE code edits of AI-generated code sourced from 1K+ developers.
First, we use \dataset to analyze when, why, and how AI-generated code is edited.
Developers invest little effort in editing code and abandon AI-generated code quickly, thereby motivating the need for LLMs to understand developer intents through their edits.
Therefore, we next use \dataset to understand to what extent LLMs could infer developer edits.
Predicting code edits is a challenging task for both frontier and fine-tuned models. 
Frontier models generate edits that best reflect developer intents, but are prone to over-editing and reverting developer-written edits.
Fine-tuning LLMs can improve their ability to predict edits that are more similar to developers' as well as better reflect developer intent, but do not fully bridge the gap to frontier models.
Together, our results demonstrate the unique insights on developer and LLM behavior that can be obtained through studying realistic code edit data to promote more better developer-AI interactions.

\end{abstract}

\begin{figure*}[t]
\centering
\includegraphics[trim=0 225 0 0, clip, width=\linewidth, keepaspectratio, page=1]{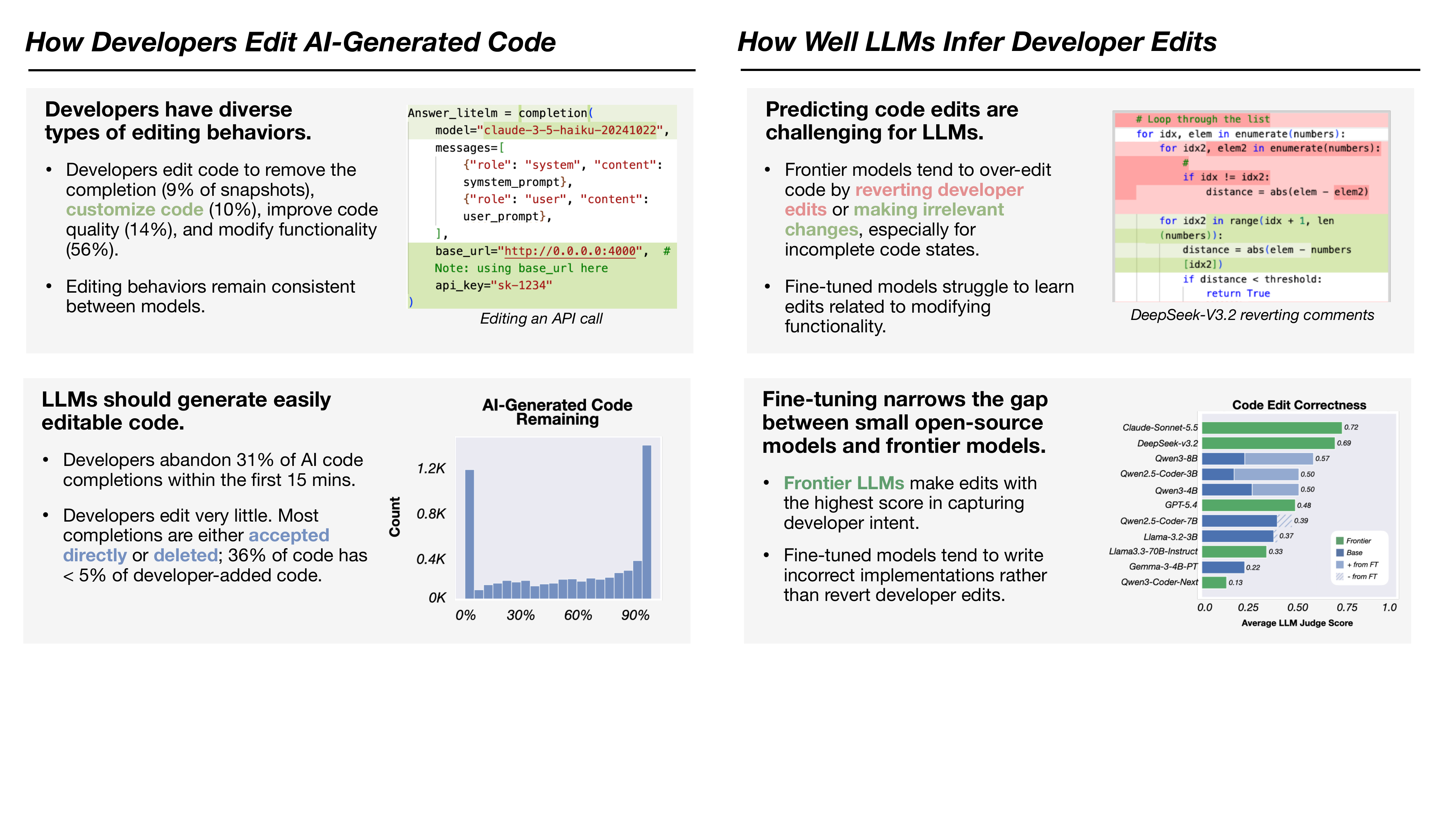}
\caption{
\textbf{Developer and LLM code editing behaviors on \dataset.}
Using \dataset, we obtain unique insights on developer and LLM behavior on code edits that is difficult to obtain from other code editing datasets.
We report our findings on how developers edit AI-generated code (\S\ref{sec:data-analysis}, left) and how well LLMs infer developer edits (\S\ref{sec:experiments}, right).
}
\label{fig:findings}
\end{figure*}

\section{Introduction}
\label{sec:introduction}

Extensive adoption of AI programming assistants has made them a common tool for software developers~\cite{anthropic2026claude, cursor2026cursor, github2026github}.
Yet, AI-generated code is imperfect, as generated code snippets can contain bugs or do not match a specific coding style~\cite{liang2024large}, and thus may require manual alterations to the generation (\ie{} code edits).
In practice, developers manually edit code generations from natural language: 44\% agent-written code is retained~\cite{baumann2026swe} and 60\% of code generated through prompting requires manual alterations~\cite{nam2025prompting}.
For coding agents, code-based interactions are also associated with higher work satisfaction and increased perceived productivity compared to natural language interactions, especially for tasks that require more fine-grained alterations \cite{brandebusemeyer2026developers}.

However, little is known about how developers manually modify AI-generated code, and to what extent LLMs are able to infer implicit developer intent.
Both are crucial for designing and building usable AI-developer interactions.
The former provides an account of how AI programming assistants fail to account for developer needs, while the latter establishes how well current models recognize those needs from code edits.
Prior research on the code editing ability of LLMs does so using publicly available code sources like GitHub data \cite{wei2024coeditor, github2025evolving, lu2025next} or competitive programming platform submissions \cite{chae2024coffee}.
However, these sources lack important code edit context, such as intermediate edits and the temporal ordering of actions~\cite{github2025evolving}.
To our knowledge, no public dataset captures real-world developer edit trajectories of AI-generated code.

To address this gap, we introduce \dataset (\textbf{\underline{D}}eveloper \textbf{\underline{E}}dits of \textbf{\underline{Co}}de \textbf{\underline{D}}atas\textbf{\underline{e}}t), a dataset of \datasetSize real-world in-IDE code edits of AI-generated code in Python, TypeScript, and JavaScript, sourced from 1K+ developers.\footnote{We will publicly release the code and dataset upon acceptance.}
This data is derived from a Visual Studio Code (VS Code) extension that enabled access to a variety of LLMs by providing AI code completions from a developer's code file context.
By extracting and tracking localized edits to actual accepted AI completions, \dataset diverges from prior code editing datasets by serving as a source of \emph{realistic}, \emph{diverse}, and \emph{high-quality} edits of AI-generated code.

In this work, we use \dataset to study how developers edit AI-generated code, and how well LLMs can infer implicit developer intent from such data (Figure~\ref{fig:findings}).
First, we use the dataset to analyze when, why, and how AI-generated code is edited (\S\ref{sec:data-analysis}).
In general, developers expect AI-generated code to be quickly usable.
They invest little effort in editing completions and abandon AI-generated code quickly, motivating the need for models to understand code edits.
We therefore then use \dataset to understand how well LLMs predict final code state from a single edit, which is challenging for both frontier and fine-tuned models.
Frontier models generate edits that best reflect developer intents, but are prone to over-editing code and often revert developer-written edits.
Fine-tuning can help smaller models predict edits that are more similar to developers' and that better capture developer intents, but they still lag behind frontier models.
Together, our results demonstrate the unique insights into developer and LLM behavior that can be obtained through studying realistic code edit data to promote better developer-AI interactions.
\begin{table*}[t]
\centering
\small
\begin{tabular}{p{3.25cm}|p{2.25cm}p{2.4cm}p{2.6cm}p{1.2cm}p{1.7cm}}
\toprule
\textbf{Dataset} & \textbf{Edit Setting} & \textbf{Real-World} & \textbf{Source} & \textbf{\# Prob.} & \textbf{Prog. Lang.} \\
\midrule
OCEDataFT \newline \citep{zhang2025generating} & Instructed edit & \xmark & LLM & 20.0K & \pythonicon \\
\addlinespace[3pt]
CanItEdit \newline \citep{cassano2024can} & Instructed edit & \xmark & Laboratory & 210 & \pythonicon \\
\addlinespace[3pt]
EditEval \newline \citep{li2024instructcoder} & Instructed edit & \xmark & LLM & 113.7K & \pythonicon \\
\addlinespace[3pt]
Coffee \newline \citep{chae2024coffee} & Instructed edit & \xmark & Code IDE & 44.8K & \pythonicon \\
\addlinespace[3pt]
PyCommit \newline \citep{wei2024coeditor} & Code edit & \cmark & GitHub & 1K & \pythonicon \\
\addlinespace[3pt]
Pencil Code \newline \citep{ross2025modeling} & Code edit & \xmark & Block-based IDE & 3.5M & \coffeescripticon \javascripticon \htmlicon \\
\midrule
\dataset (ours) & Code edit & \cmark & Code IDE & 53.6K & \pythonicon \typescripticon \javascripticon \\
\bottomrule
\end{tabular}
\caption{\textbf{\dataset is the only dataset of real-world manual code edits by developers.} We compare \dataset to other code editing datasets by edit setting (\emph{instructed edits} via natural language prompts or manual \emph{code edits}), focus on real-world edits and development tasks, source, problem count, and languages (Python \pythonicon, JavaScript \javascripticon, TypeScript \typescripticon, CoffeeScript \coffeescripticon, HTML/CSS \htmlicon).}
\label{tab:dataset-comparison}
\end{table*}

\section{Related Work}
\label{sec:related-work}

\paragraph{Developer Studies of Editing AI-Generated Code.}
Both natural language and code-based interactions with AI programming assistants are valuable.
While natural language interactions improve developer productivity more, code edits result in higher work satisfaction and perceived productivity, and are cognitively less demanding~\citep{liang2024large, brandebusemeyer2026developers}.
Building on this literature, a line of research has investigated how developers edit AI-generated code from natural language, which is most related to this work.
\citet{nam2025prompting} collected and analyzed interaction logs from 1K Google developers using AI programming assistants to edit code in natural language.
They found that after accepting an edit from the assistant, developers directly edited the code 62\% of the time, while in 12\% of cases developers re-prompted the tool.
Meanwhile, \citet{baumann2026swe} collected a dataset of 18K coding agent sessions.
Their analysis revealed that developers engaged in three types of interaction modes with coding agents---human-only coding (25\% of sessions), collaborative coding (34\% of sessions), and vibe coding (41\% of sessions)---which resulted in 55\% of code being agent-written.
We complement this literature by focusing on how developers express intents implicitly via code edits, rather than explicitly via natural language instructions.

\begin{figure*}[t]
    \centering
    \includegraphics[width=0.9\linewidth, trim=0 650 300 0, clip, keepaspectratio, page=1]{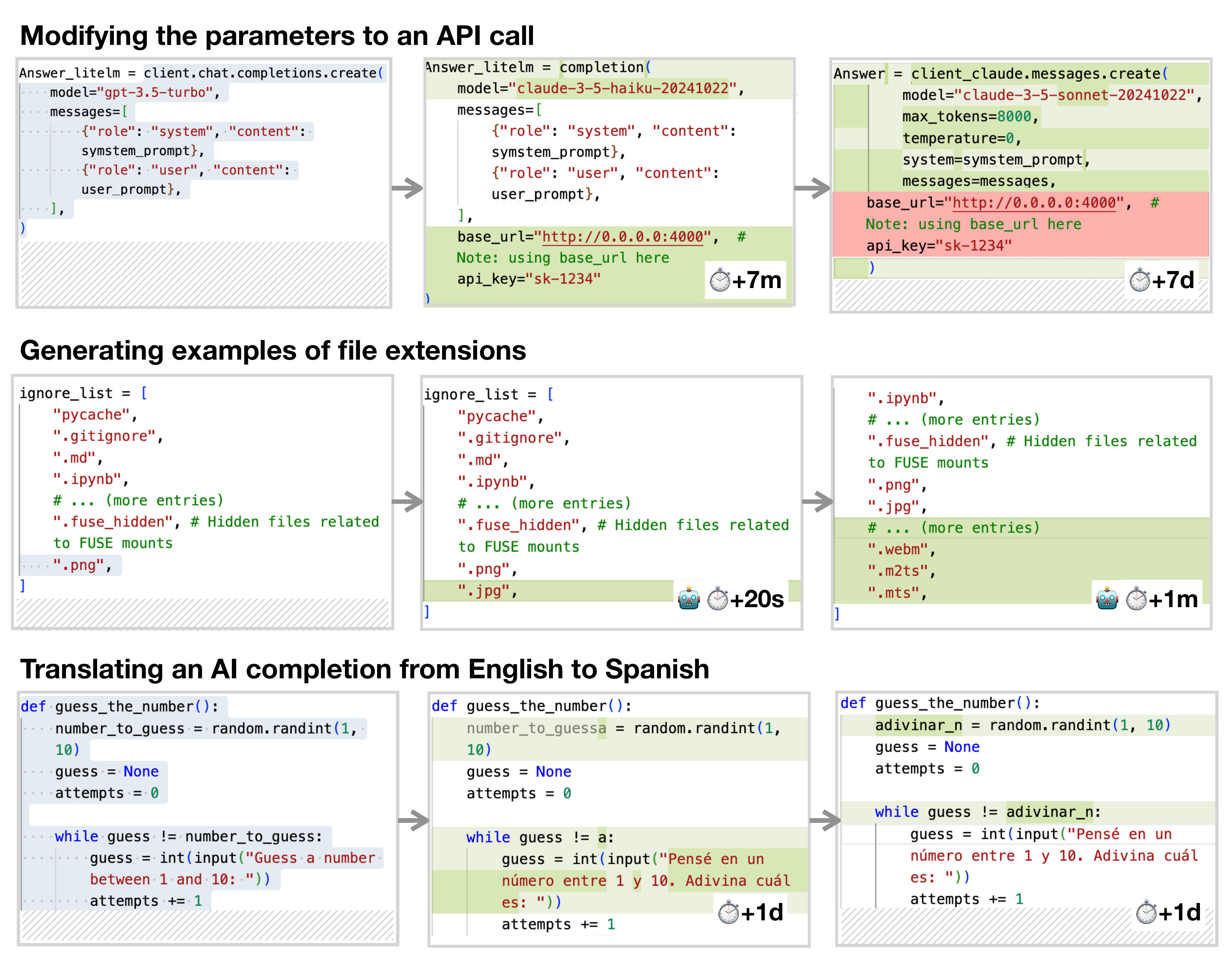}
    \caption{\textbf{\dataset captures diverse code editing behavior.}
    We show three examples of developer code edit trajectories in \dataset with the AI completion (gray), its subsequent removal (red) and addition (green) edits, the time between accepting the AI completion and making the edit (\clockIcon), and whether the edit was AI-generated (\robotIcon).
    }
    \label{fig:example-trajectories}
    \vspace{-5pt}
\end{figure*}

\paragraph{Predicting Code Edits.}
Several works evaluate LLMs' abilities to predict edits from natural language instructions in laboratory~\cite{cassano2024can, muennighoff2023octopack} and real-world~\cite{chi2025editbench} settings, and found that LLM performance differs by model size and openness.
Approaches for modeling code edits are fill-in-the-middle LLMs~\cite{fried2023incoder, roziere2023code, guo2024deepseek} 
or directly modeling sequences of code changes~\cite{zhang2022coditt5, gupta2023grace, li2023codeeditor, wei2024coeditor, github2026copilot}.
Several works have developed code edit datasets with GitHub commits~\cite{wei2024coeditor, lu2025next}, competitive programming problem submissions~\cite{chae2024coffee, shypula2024learning}, student programming problem submissions~\cite{ross2025modeling}, and synthetic data~\cite{zhang2025generating, li2024instructcoder}.
Most related to \dataset is Overwatch~\cite{zhang2022overwatch}, a private dataset of code edits from 12 Microsoft developers.

\dataset differs from existing datasets by being the first public dataset to obtain code edits from real-world software developer contexts at scale (see Table~\ref{tab:dataset-comparison}).
Existing work either focuses on the task of edits in the form of instructions, or obtains code edits via simulations or under constrained development settings.
These data sources pose limitations, as our experiments indicate that models struggle to predict the final state of edits with unfinished code states common in code editing (\S\ref{sec:experiments-results}).

\section{\dataset: Developer Edits of Code Dataset}
\label{sec:dataset}

\dataset contains \datasetSize real-world in-IDE developer code edits of AI-generated code in Python, TypeScript, and JavaScript.
We formulate the code editing problem for \dataset as follows.
Given an initial version of an accepted AI code completion $y_0$ within a code file, a developer $d$ will edit the code snippet $y_0$ at time $t$, producing a new \emph{snapshot} of the code snippet $y_t$.
Over time, this produces a \emph{trajectory} of edits made by developer $d$ on the original completion: $Y^d=[y_0^d, y_1^d, y_2^d,...,y_t^d]$.
See ~\cref{fig:example-trajectories} for examples.

\subsection{Dataset Construction}
\label{sec:dataset-construction}
To extract code edits for \dataset, we develop a pipeline that follows a four-step process (see~\cref{sec:appendix-dataset-construction} for full details):

\paragraph{\textnormal{\textbf{Step 1: Obtaining Code Files.}}}
\dataset is constructed using data from a popular VS Code extension for developers to access a variety of frontier models for coding assistance.
The study is approved by a Institutional Review Board (IRB).
In addition to AI code completions, the extension saves the current working code file each time a developer pauses activity in the IDE for one second, yielding a rich source of code edit data.
The pipeline begins with a pool of 4,110 developers who accepted AI code completions on the extension.
We collect their code files and identify those likely containing edits of $y_0$ through a rough file header match based on line overlap between the code file $y_0$ and all other code files associated with $d$.
We remove redundant data by discarding snapshots within 10 seconds of the previous one.

\paragraph{Step 2: Code Edit Extraction.}
After identifying candidate code files that could contain snapshots of $y_0$, we extract the code snippet relevant to the edit from each file.
A key challenge is isolating the lines of code directly relevant to an edit, since the data quality of the code edits is more important than the data quantity for model training~\cite{github2025evolving}.
Our approach for accurately extracting code edits relies on \texttt{\small{git diff -\,-histogram}} to identify differences in the source code~\cite{nugroho2020different}.
For each file containing an edit snapshot $y_t$, we generate a ``diff'' between $y_t$ and $y_{t-1}$ to locate and map the lines of code that were edited, relative to the original snapshot $y_0$.
The \texttt{git diff} filtering step reduces the dataset from 6.5M lines of code overall to 772.2K lines of code directly relevant to code edits of AI completions.
An example of an extraction is shown in~\cref{fig:example-extraction}. 

\paragraph{Step 3: Data Cleaning \& Validation.}
After obtaining the initial trajectories, we clean each trajectory $Y^d$ by removing malformed data and validating each $y_t$'s relevance to $y_0$ with CodeBERTScore~\cite{zhou2023codebertscore} threshold of 0.68, which was selected based on preliminary analysis.
Data cleaning reduces the dataset from 78K before-after pairs from 4K developers to 56K before-after pairs from 1K developers. 
Given the importance of data quality for code editing~\cite{github2025evolving}, we validate the pipeline by annotating a sample of 125 before-after pairs that were the first and last snapshots from a sample of edit trajectories.

\paragraph{Step 4: PII Redaction.}
We reduce the likelihood that \dataset contains any personally identifiable information (PII) by reusing the PII redaction pipeline used by Starcoder and the Stack~\cite{li2023starcoder, lozhkov2024starcoder}, which relies on an LLM trained on PII contained in code files.
We additionally apply OpenAI Privacy Filter~\cite{openai2026privacy}, a general-purpose PII redaction model.

\begin{figure*}[t]
    \centering
    \begin{subfigure}[b]{0.32\textwidth}
        \centering
        \includegraphics[width=\textwidth, trim=0 475 1200 0, clip, keepaspectratio, page=1]{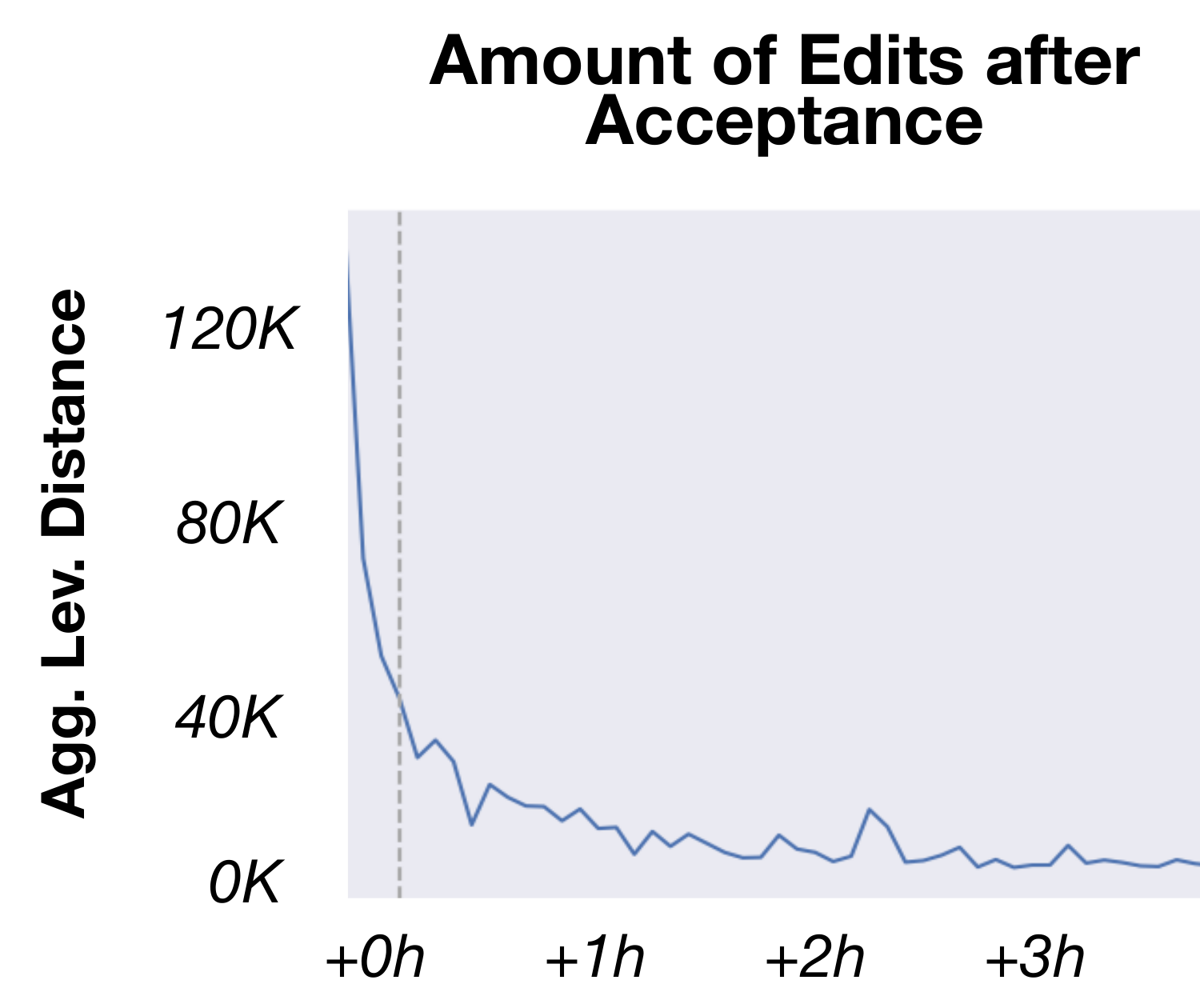}
        \phantomcaption{}
        \label{fig:when-edits}
    \end{subfigure}
    \begin{subfigure}[b]{0.32\textwidth}
        \centering
        \includegraphics[width=\textwidth, trim=0 475 1200 0, clip, keepaspectratio, page=2]{figures/analysis-results.pdf}
        \phantomcaption{}
        \label{fig:code-remaining}
    \end{subfigure}
    \begin{subfigure}[b]{0.32\textwidth}
        \centering
        \includegraphics[width=\textwidth, trim=0 475 1200 0, clip, keepaspectratio, page=3]{figures/analysis-results.pdf}
        \phantomcaption{}
        \label{fig:code-added}
    \end{subfigure}
    \caption{\textbf{Developer editing behavior after accepting AI-generated code.}
    \emph{Left:} The number of changes made to AI completions drops after the first 15 minutes (vertical gray dotted line).
    \emph{Middle:} Developers often completely remove the completion or leave it fully intact.
    \emph{Right:} Developers rarely add new code to AI-generated code.
    }
    \label{fig:four_panel}
\end{figure*}

\subsection{Dataset Statistics}
\label{sec:dataset-statistics}

\dataset contains $5,831$ trajectories from $1,141$ developers corresponding to $53,614$ before-after code pairs from 20 different LLMs (see full details in \cref{sec:appendix-dataset-statistics}).
Before-after code pairs are written in Python ($40,390$), JavaScript ($7,993$), and TypeScript ($5,231$), spanning a diverse set of software development task contexts: data analysis and visualization ($1,285$), web development ($1,179$), machine learning ($901$), web scraping ($706$), user interfaces ($609$), file processing ($451$), backend development ($371$), game development ($204$), and testing ($125$).
Edit trajectories in \dataset last for a median duration of $49.7$ minutes (min: $1$ second, max: $283.5$ days) and the number of snapshots per trajectory has a median of $4$ snapshots (min: $1$ snapshot, max: $424$ snapshots).
\section{How Do Developers Edit AI-Generated Code? (RQ1)}
\label{sec:data-analysis}
We use \dataset to study how developers edit AI-generated code with the following metrics.
The purpose of this investigation is to use the insights on developer behavior to understand where LLM generations fail to address developer needs.

\subsection{Metrics}
\label{sec:data-analysis-methodology}

We use the following metrics.
For more details, see~\cref{sec:appendix-developer-metrics}. 
\begin{itemize}[leftmargin=*]

\item \textbf{Amount of AI-generated code remaining:} The proportion of an AI completion remaining at the end of a trajectory based on the Levenshtein removal and replacement operations.

\item \textbf{Amount of developer-added code:} The proportion of code at the end that was added by the developer and was not part of the original completion using the Levenshtein addition operations.

\item \textbf{Total amount of code edits:} The total amount of edits made by all developers at a specific time step, computed by aggregating Levenshtein distances made within the same time period.
\end{itemize}

\subsection{Results}
\label{sec:data-analysis-results}

\paragraph{Finding 1: Developers' editing behaviors are diverse.}
We identify four types of code edits by analyzing a sample of edit trajectories and using an LLM judge to classify snapshots (see ~\cref{sec:appendix-developer-code-edits}).
The distribution of edit types is broadly similar across models.
Though differences are statistically significant ($\chi^2=320.7$, $p<0.001$), the effect size is small (Cram\'{e}r's $V=0.05$;~\citealp{cohen2013statistical}).
The edit types are:
\begin{itemize}[leftmargin=*]

\item \textbf{Customizing code}: Code edits to adjust the AI completion to better align with the developer's intent without significantly changing the code's functionality, accounting for 10\% of edit snapshots. Examples include changing parameter, variable, and literal values (\egCustom changing the text passed to \texttt{print()}) and renaming variables.
    
\item \textbf{Improving code quality}: Code edits to improve the quality of the AI-generated code (\eg{} fixing syntax errors or adding comments), constituting 14\% of edit snapshots.
    
\item \textbf{Changing code functionality}: Code edits to change the AI-generated code's behavior, such as adding new methods or changing logic, accounting for 56\% of edit snapshots.

\item \textbf{Removing}: Code edits with the intent of removing the AI completion from the code and starting from the original state (\egCustom commenting out the completion), representing 9\% of edit snapshots.
\end{itemize}

\begin{figure}[t]
\centering
\includegraphics[trim=0 325 975 0, clip, width=0.8\linewidth, keepaspectratio, page=4]{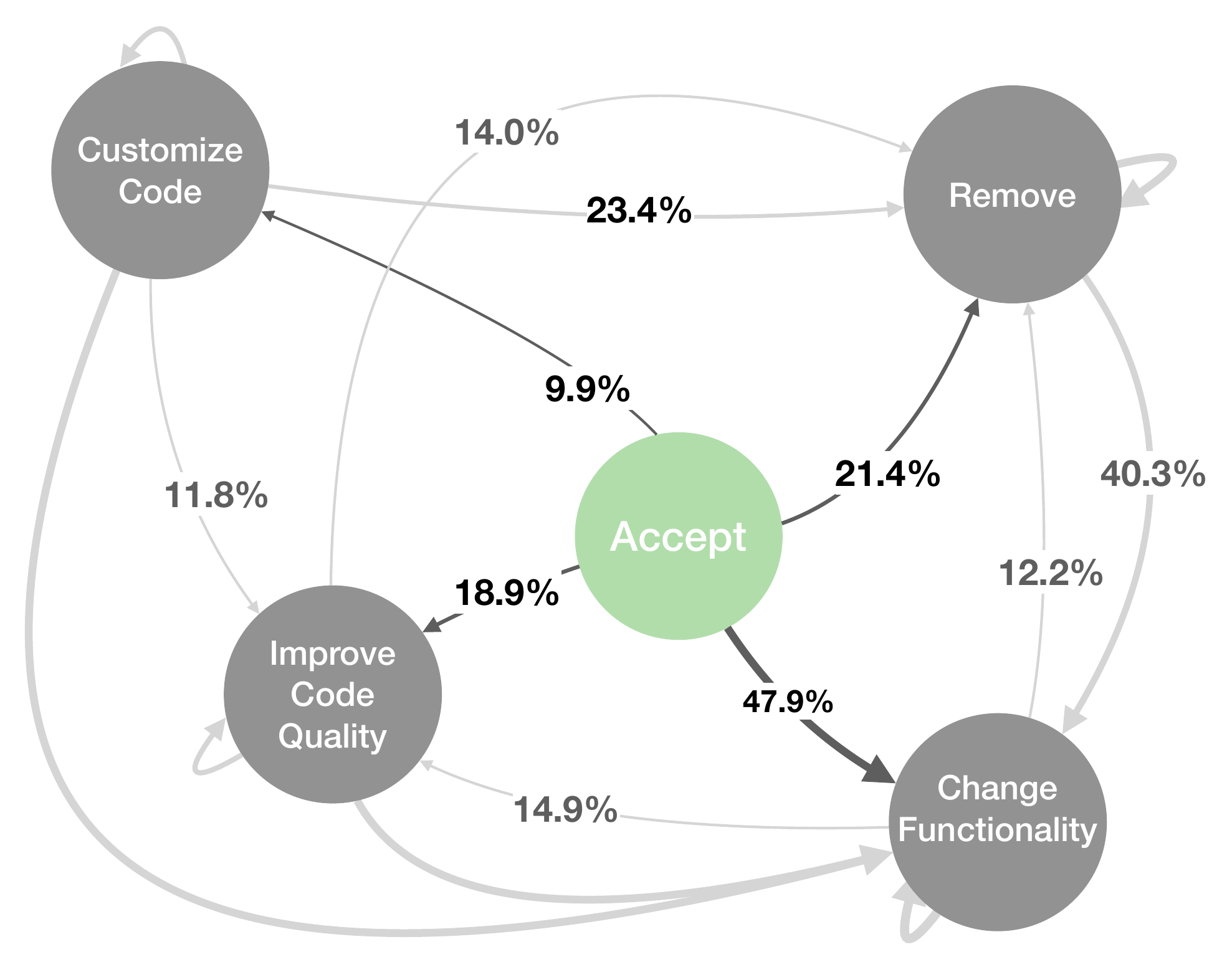}
\caption{
\textbf{Editing behavior after accepting AI code completions.}
We show the first (black arrow) and second (gray arrow) edits after accepting a completion.
We present second edits that are frequent ($\geq$ 10\%).
}
\label{fig:edit-trajectories}
\end{figure}

\paragraph{Finding 2: Developers expect AI-generated code to be usable almost immediately; otherwise, they delete it.}
Most edits occur very shortly after acceptance, as there is a steep drop in the total number of edits 15 minutes after acceptance (\cref{fig:when-edits}).
In the first 15 minutes, a large portion of AI completions are abandoned within the trajectory (31\%).
In fact, 21\% of the first edits remove the completion (\cref{fig:edit-trajectories}).
In some cases, developers attempt to adapt the code before removing it.
Removal is more common after an initial attempt to customize the code (23\%) than after other types of edits, indicating developers are willing to abandon code when it did not meet their use case.

Developers also do not expend much effort to adapt AI-generated code.
Developers either accept AI completions with few modifications or completely remove them, since the amount of code remaining of the AI code completion is bimodal (\cref{fig:code-remaining}).
Developer additions are also limited in many cases, as 36\% of the final code contains less than 5\% of developer-added code (\cref{fig:code-added}).
This behavior is generally consistent across models.
Kruskal-Wallis tests~\cite{mckight2010kruskal} with Benjamini-Hochberg correction also shows significant differences in remaining AI-generated code ($H=31.5$, $p=0.04$) and developer-added code ($H=57.3$, $p<0.001$), but with small effect sizes ($\eta^2=0.002$ and $0.007$)~\cite{fiel2026effect}.

\section{How Well Do LLMs Infer Developer Edits? (RQ2)}
\label{sec:experiments}
Developers invest little effort to edit misaligned code and abandon AI-generated code quickly, resulting in nearly a third of AI code completions being discarded (\S\ref{sec:data-analysis-results}).
Thus, having models that infer developer intent based on their code edits is imperative for improved developer-AI interactions.
In this section, we explore this idea by evaluating how well LLMs infer developer edits.

\subsection{Experimental Setup}
\label{sec:experiments-setup}

\paragraph{Task.}
In order to simulate the real-world setting of a developer editing AI-generated code, we formulate the task as follows: given an AI completion $y_0$ and the first edit snapshot $y_1$, predict $y_t$, the final code state.

\paragraph{Frontier Models.}
To evaluate frontier LLMs' existing abilities to reason about code edits, we experiment with few-shot in-context learning with frontier models.
We prompt the following models: \modelName{Llama-3.3-70B}~\cite{dubey2024llama3}, \modelName{Qwen3-Coder-Next} ~\cite{cao2026qwen3codernext}, \modelName{GPT-5.2}~\cite{singh2025gpt5systemcard}, \modelName{Claude-Sonnet-5.5} ~\cite{anthropic2026claudesonnet55}, and \modelName{DeepSeek-v3.2}~\cite{deepseek2024v3}.
We prompt each model with three before-after pairs from \dataset; supply the code prefix, AI completion, and code suffix; the developer's edit; and instruct the LLM to generate the completion's final state after editing.

\paragraph{Fine-Tuned Models.}
To study LLMs' abilities to learn from code edits, we fine-tune six models on \dataset: \modelName{Qwen2.5-Coder-3B}~\cite{hui2024qwen2}, \modelName{Qwen2.5-Coder-7B}, \modelName{Qwen3-4B}~\cite{yang2025qwen3}, \modelName{Qwen3-8B}, \modelName{Llama3.2-3B}~\cite{dubey2024llama3}, and \modelName{Gemma3-4B-IT}~\cite{team2025gemma}.
To understand whether fine-tuned models improved through fine-tuning, we compare them to their base variants using the same few-shot setup as the frontier models.
Models are trained using LoRA~\cite{hu2022lora}.
Full experimental details are in~\cref{sec:appendix-modeling-setup}.

Since code generation benefits from sampling~\cite{li2022competition, shypula2024learning}, we follow~\citet{shypula2024learning} and use beam search with a beam width of 3 and temperature of 0.7~\citep{graves2012sequence}.
We set the maximum new tokens for all models to be the 95th percentile of the training set's final edit lengths.

\paragraph{Dataset Variants.}
We create a subset of \dataset known as \bigDataset.
This dataset corresponds to train, development, and test sets of real-world code edits from 80\%, 10\%, and 10\% of the developers respectively (randomly split) from \dataset to prevent data leakage.
The purpose of this data is to facilitate whether LLMs could learn and understand developer edits at a large scale.

To complement the insights from \bigDataset and understand what the models learned, we also prepare \smallDataset, a set of 15 synthetic Python code editing problems.
This dataset is developed by two study authors who independently created and validated the problems. 
Editing problems are grounded in examples from \dataset and evenly distributed across the three edit types identified in \S\ref{sec:data-analysis-results}.
We design each problem to make edit correctness easier to assess.

\begin{table}[t]
\centering
\small
\setlength{\tabcolsep}{2.5pt}

\begin{tabular}{l|ccc|c}
\toprule
& \multicolumn{3}{c|}{\textbf{\sc{Real}}}
& \textbf{\sc{Synth.}} \\
\textbf{Edit type}
& \textbf{Lev.}
& \textbf{C-BERT}
& \textbf{C-BLEU}
& \textbf{Lev.} \\
\midrule
Customizing code
& 0.35 & 0.71 & 0.40 & 0.51 \\
+\bigDataset
& \metricDelta{green}{+0.25}
& \metricDelta{green}{+0.17}
& \metricDelta{green}{+0.14}
& \metricDelta{green}{+0.32} \\
\midrule
Improving code qual.
& 0.31 & 0.69 & 0.42 & 0.45 \\
+\bigDataset
& \metricDelta{green}{+0.18}
& \metricDelta{green}{+0.16}
& \metricDelta{green}{+0.13}
& \metricDelta{green}{+0.30} \\
\midrule
Changing code func.
& 0.27 & 0.66 & 0.34 & 0.47 \\
+\bigDataset
& \metricDelta{green}{+0.12}
& \metricDelta{green}{+0.12}
& \metricDelta{green}{+0.06}
& \metricDelta{green}{+0.07} \\
\bottomrule
\end{tabular}
\caption{\textbf{Edit similarity across edit types between pre-/post- fine-tuning on \bigDataset.}
Models experienced the least gains for edits involving changing code functionality after fine-tuning.
We report normalized Levenshtein, CodeBERTScore (C-BERT), and CodeBLEU (C-BLEU) scores averaged across models, followed by the corresponding changes after fine-tuning on \bigDataset (+\dataset, \metricDelta{green}{+XX}). Higher is better.
}
\label{tab:modeling-results-by-edit-type}
\end{table}

\paragraph{Metrics}
We evaluate LLMs with text similarity metrics, as there are no tests with \dataset's code files.
This precludes functional correctness measures like pass@k~\citep{chen2021evaluating}.
We report normalized Levenshtein similarity---following prior work~\citep{wei2024coeditor, lu2025next}---CodeBERTScore~\citep{zhou2023codebertscore}, and CodeBLEU~\citep{ren2020codebleu}.
We interpret higher scores as closer agreement with observed developer edits, rather than definitive evidence of better edit quality or greater understanding of developer intent.

For \smallDataset, we evaluate correctness using problem-specific rubrics, scoring each item as 0 (reverting developer edits), 1 (retaining developer edits without making additional changes), 2 (partially correct), and 3 (fully correct) with Krippendorff's $\alpha = 0.84$~\cite{krippendorff1989content} between an LLM judge and one human rater.
We use this metric as an exploratory measure of how well models infer nuanced developer intent from code edits.
We also manually annotate lines that should remain unchanged to calculate the number of lines the LLM over-edited.

\subsection{Results}
\label{sec:experiments-results}

\begin{figure}[t]
\centering
\includegraphics[trim=0 525 1250 0, clip, width=\linewidth, keepaspectratio, page=1]{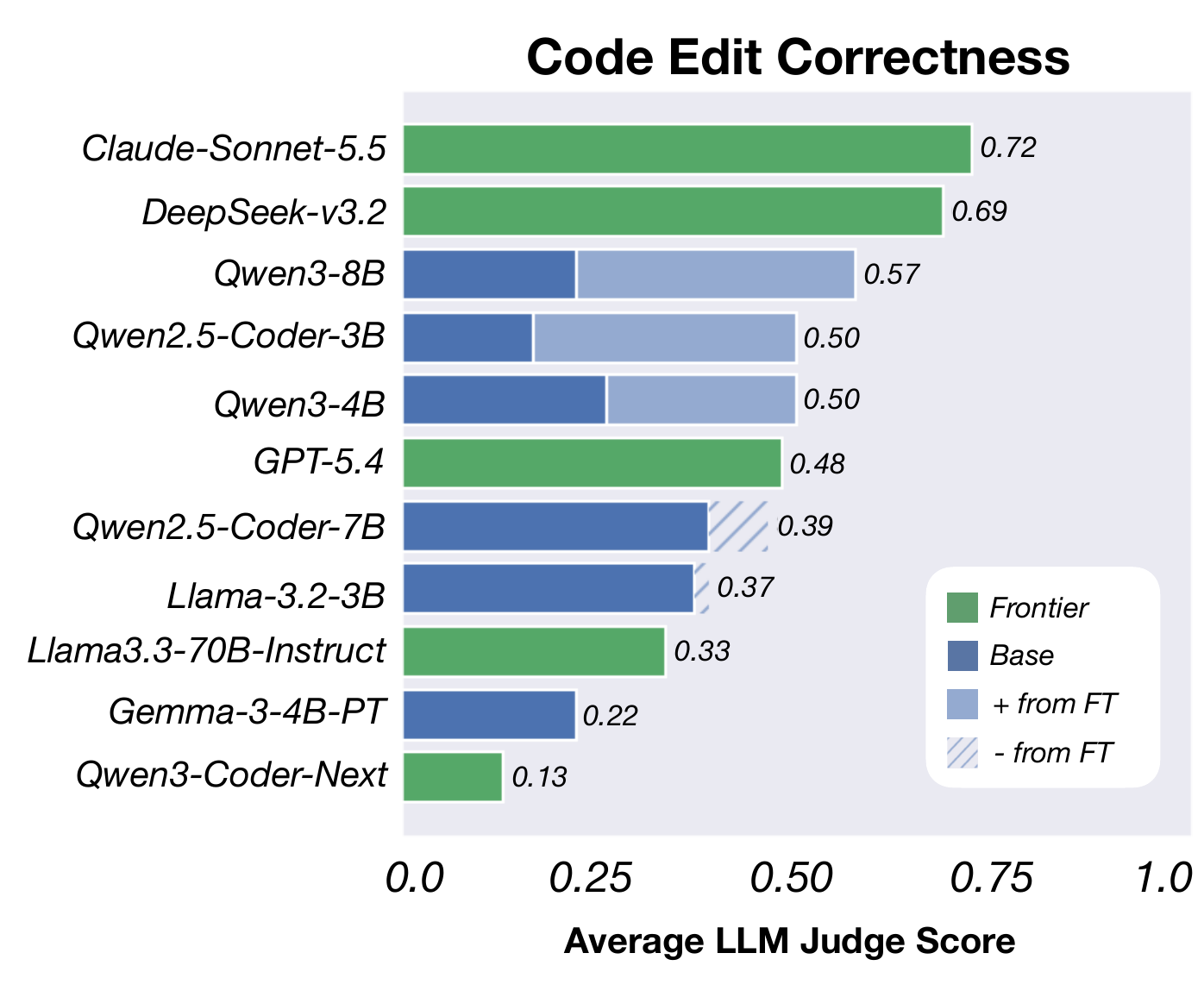}
\caption{
\textbf{Average edit correctness on \smallDataset code editing problems.}
Frontier models (\modelName{Claude-Sonnet-5.5}, \modelName{DeepSeek-v3.2}) scored highest in producing correct edits that reflected developer intent.
}
\label{fig:edit-correctness}
\end{figure}

\begin{table}[t]
\centering
\small
\setlength{\tabcolsep}{3pt}

\begin{tabular}{l|ccc|c}
\toprule
& \multicolumn{3}{c|}{\textbf{\sc{Real}}}
& \textbf{\sc{Synth.}} \\
\textbf{Model}
& \textbf{Lev.}
& \textbf{C-BERT}
& \textbf{C-BLEU}
& \textbf{Lev.} \\
\midrule
\modelName{Qwen2.5-Coder-3B}
& 0.29 & 0.72 & 0.36 & 0.32 \\
+\bigDataset
& \metricDelta{green}{+0.18}
& \metricDelta{green}{+0.10}
& \metricDelta{green}{+0.08}
& \metricDelta{green}{+0.38} \\
\addlinespace[3pt]
\modelName{Qwen2.5-Coder-7B}
& 0.36 & 0.76 & 0.41 & 0.77 \\
+\bigDataset
& \metricDelta{green}{+0.14}
& \metricDelta{green}{+0.06}
& \metricDelta{green}{+0.05}
& \metricDelta{gray}{+0.00} \\
\midrule
\modelName{Qwen3-4B}
& 0.34 & 0.72 & 0.38 & 0.32 \\
+\bigDataset
& \metricDelta{green}{+0.14}
& \metricDelta{green}{+0.09}
& \metricDelta{green}{+0.06}
& \metricDelta{green}{+0.41} \\
\addlinespace[3pt]
\modelName{Qwen3-8B}
& 0.47 & 0.80 & 0.44 & 0.61 \\
+\bigDataset
& \metricDelta{green}{+0.02}
& \metricDelta{green}{+0.01}
& \metricDelta{green}{+0.02}
& \metricDelta{green}{+0.11} \\
\midrule
\modelName{Llama3.2-3B}
& 0.24 & 0.65 & 0.30 & 0.31 \\
+\bigDataset
& \metricDelta{green}{+0.24}
& \metricDelta{green}{+0.16}
& \metricDelta{green}{+0.15}
& \metricDelta{green}{+0.29} \\
\midrule
\modelName{Gemma-3-4B-PT}
& 0.16 & 0.45 & 0.23 & 0.53 \\
+\bigDataset
& \metricDelta{green}{+0.27}
& \metricDelta{green}{+0.32}
& \metricDelta{green}{+0.17}
& \metricDelta{green}{+0.20} \\
\bottomrule
\end{tabular}
\caption{\textbf{Edit similarity pre-/post- fine-tuning on \bigDataset.}
After fine-tuning, models produce more similar edits on both datasets.
We report the few-shot base model's normalized Levenshtein, CodeBERTScore (C-BERT), and CodeBLEU (C-BLEU) scores followed by the change after fine-tuning on \bigDataset (+\dataset, \metricDelta{green}{+XX}). Higher is better.
}
\label{tab:modeling-results}
\end{table}

\begin{figure*}[t]
\centering
\includegraphics[trim=0 600 300 0, clip, width=\linewidth, keepaspectratio, page=1]{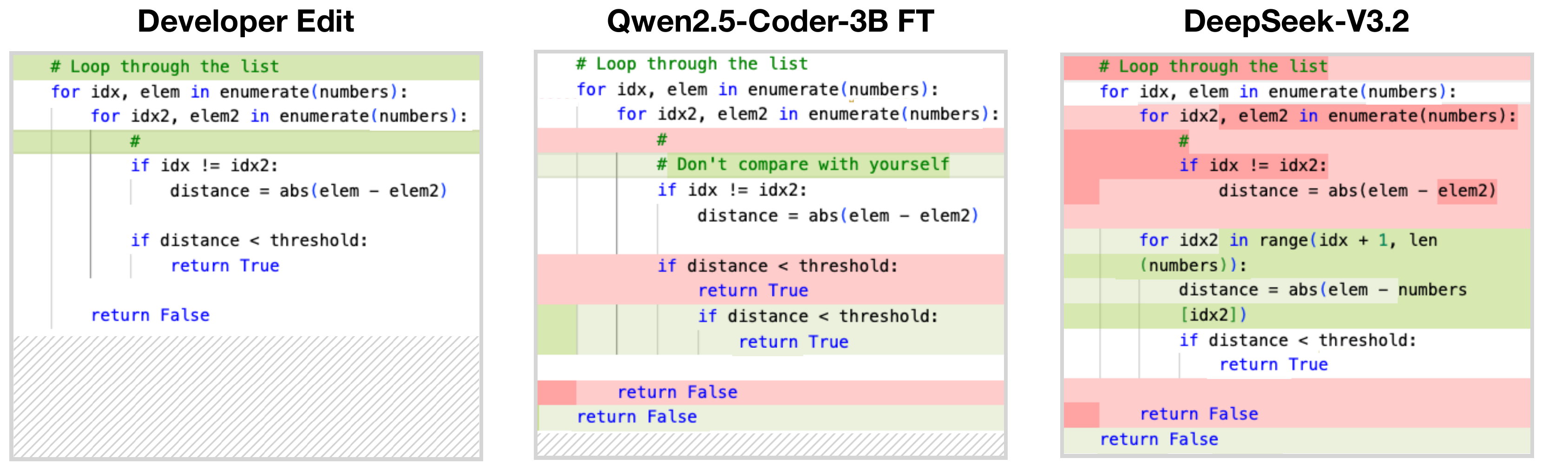}
\caption{
\textbf{Example of frontier versus fine-tuned models on \smallDataset code editing problems.}
We present an editing problem where the developer intends to add a comment.
\modelName{DeepSeekV3.2} over-edited by making significant changes to how the code was written, reverting the developer's edit.
}
\label{fig:over-editing}
\end{figure*}

\paragraph{\textbf{Finding 3: Predicting code edits is a challenging task for LLMs.}}
The frontier models \modelName{Claude-Sonnet-5.5} and \modelName{DeepSeek-v3.2} score the highest on \smallDataset (Figure~\ref{fig:edit-correctness}) with an average score of 0.71 and 0.69 respectively.
However, they have room to improve, as the maximum score is 1.0.
Most (8 of 12) their errors occurred on edits resulting in incomplete code states, where the model reverted the developer's edit in lieu of writing the model's preferred implementation, such as removing a comment after a developer added it (Figure \ref{fig:over-editing}) or deleting a developer's starting code.
Reverting a developer's edits holds for frontier models broadly: 71\% of model errors are due to undoing developer edits to prefer the original completion, ignoring explicit developer feedback.
When the edit resulted in a more completed code state (4 of 12 errors), the frontier models failed to continue the edit.

One reason why code editing is challenging is knowing where to edit.
Frontier models tend to make unnecessary edits to other code regions, sometimes at the expense of the developer's intent (example in~\cref{fig:over-editing}).
On average, frontier models edited 3.5 lines of code unnecessarily, whereas fine-tuned models over-edited less than half as many (1.5 lines on average).
Through fine-tuning on \bigDataset, models tend to over-trust developer edits and refrain from editing code that developers changed.
In fact, 26\% of errors involve not changing the code after the developer edited it.

Another reason why code editing is challenging is learning \emph{specific} types of edits.
For both datasets, LLMs had the least improvement on edits related to modifying functionality (\cref{tab:modeling-results-by-edit-type}), despite these edits being the most prevalent, representing 56\% of all edit snapshots (\S\ref{sec:data-analysis-results}).
This may be because this type of edit requires a semantic understanding of code, which is harder for LLMs to learn than code syntax~\citep{zhao2026syntax, kumar2026syntax}.

\paragraph{Fine-tuning on \dataset narrows the gap between small open-source models and frontier models.}
Fine-tuning improves LLMs' ability to produce more similar edits as developers (Table~\ref{tab:modeling-results}), as well as edits that better reflect developer intent (Figure~\ref{fig:edit-correctness}).
The best fine-tuned models, \modelName{Qwen3-8B} and \modelName{Qwen2.5-Coder-3B} score 0.53 and 0.51 on edit correctness, indicating that fine-tuning does not fully close the gap to frontier models. 
Fine-tuned models tend to fail due to writing incorrect implementations (10 of 20 errors) when the code is an incomplete state, rather than reverting to the completion (6 errors) or not continuing the developer's edit (4 errors).
With seemingly complete code (8 errors), fine-tuned models further modify the code to an incorrect solution (6 errors) rather than making no changes (2 errors), indicating that these models may struggle with how to change the code rather than whether to make a change at all.

\section{Conclusion}
\label{sec:discussion}
AI-generated code is imperfect, often requiring developers to make manual \emph{code edits}.
Code edits are important in for today's AI programming assistant landscape---including for coding agents---since developer's routinely make modifications to code generated from natural language.
Yet, little is known about how developers manually edit AI-generated code and to what extent LLMs can infer implicit developer intent.
We introduce \dataset, a dataset of \datasetSize developer edits of AI-generated code.
By analyzing \dataset, we find that developers quickly abandon AI code completions and invest little effort in adapting the code, motivating the need for LLMs that infer developer intent via code edits.
We explore this idea by benchmarking LLMs' capabilities to infer developer edits.
Frontier models generate edits that best reflect developer intents, but are prone to over-editing and reverting developer-written edits.
Fine-tuning LLMs can produce edits that are more similar to developers' and better reflect their intent, but do not fully bridge the gap to frontier models.
Our results demonstrate the unique insights on developer and LLM behavior that may be obtained by studying realistic code edit data for better developer-AI interactions.

\section*{Limitations}
\dataset captures edits from a fixed set of LLMs within a specific time window and only includes edits to accepted AI-generated code.
This excludes rejected completions, human-written code, and code obtained from outside the VS Code extension.
Also, the task of generating code edits is subjective since developers vary in how they write code, and there are multiple valid ways to edit a completion.
Thus, \dataset may not capture the full diversity of code edits, which can affect performance on the edit generation task and limit the ability of fine-tuned LLMs to generalize across different editing styles.
While fine-tuning LLMs on \dataset, we selected the best-performing check points based on normalized Levenshtein score.
As a result, LLMs may not improve uniformly across all evaluation metrics, especially because normalized Levenshtein distance is not perfectly correlated with all other evaluation metrics.

\section*{Ethical Considerations}
All data in \dataset are collected from developers who have explicitly opted to sharing their code data via a VS Code extension.
Before releasing the dataset, we remove personally identifiable information (PII) from the entire dataset.
More specifically, we reuse the PII redaction pipeline used by Starcoder and the Stack~\cite{li2023starcoder, lozhkov2024starcoder}.
We additionally apply OpenAI Privacy Filter~\cite{openai2026privacy}, a general-purpose PII redaction model.
The study is approved by an Institutional Review Board (IRB).

\bibliography{custom}

\appendix
\section*{Appendix}
\label{sec:appendix}

\section{\dataset: Developer Edits of Code Dataset}
\label{sec:appendix-dataset}

\begin{figure*}[t]
    \centering
    \includegraphics[width=0.9\linewidth, trim=2 550 350 0, clip, keepaspectratio, page=1]{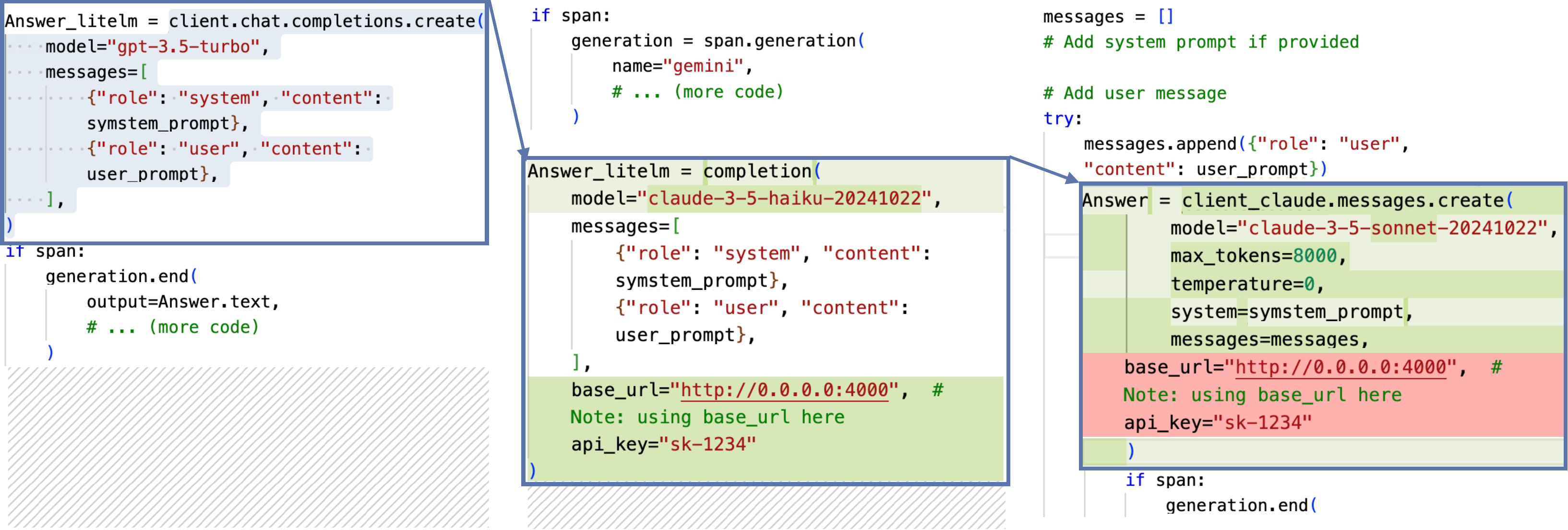}
    \caption{\textbf{Example code edit extraction from \dataset.}
    The data extraction pipeline begins with identifying an original AI completion (gray code snippet).
    It tracks what changes made to the AI completion, including additions (green) and removals (red), even when the developer changes the surrounding code.
    }
    \label{fig:example-extraction}
    \vspace{-5pt}
\end{figure*}

\subsection{Dataset Construction}
\label{sec:appendix-dataset-construction}
Below, we enumerate additional details of the data construction process of \dataset.  

\paragraph{Languages Selected.} The VS Code extension consisted of completions in several languages, but we selected all Python, Javascript, and Typescript completions as they were the three most popular languages, comprised well over half of the available completions, and were supported well by CodeBERT with only slight pre-processing.

\paragraph{Models in Edit Data.} The VS Code extension, and subsequently \dataset, includes code completions from the following models: \modelName{gpt-4o-mini-2024-07-18}, \modelName{llama-3.1-70b-instruct}, \modelName{llama-3.1-405b-instruct}, \modelName{codestral-2405}, \modelName{deepseek-coder-fim}, \modelName{gemini-1.5-flash-002}, \modelName{gemini-1.5-pro-002}, \modelName{claude-3-5-sonnet-20240620}, \modelName{gpt-4o-2024-08-06}, \modelName{gpt-4o-2024-11-20}, \modelName{qwen-2.5-coder-32b-instruct}, \modelName{claude-3-5-sonnet-20241022}, \modelName{gemini-2.0-flash-exp}, \modelName{codestral-2501}, \modelName{deepseek-coder-v3-fim}, \modelName{anonymous-titan}, \modelName{gemini-2.0-flash-001}, \modelName{gemini-2.0-pro-exp-02-05}, \modelName{anonymous-q}, and \modelName{claude-3-7-sonnet-20250219}.

\paragraph{Matching File Headers.} To identify code files that may contain edited code snippets for an initial completion, we identify the initial 'substantive' lines in each file of the developer, and compute the similarity between the originally modified file and other files. After validation by hand-annotation, we set a similarity cutoff for 0.5 for Python files, and 0.4 for Javascript and Typescript files. 

\paragraph{Removing Edits at the Ends.} Identifying the final state of an edit that was made on the very first or very last line of a file is a challenging task as it is difficult to determine programatically where the edit ends in its final state, and other unrelated edits begin. To preserve the quality of \dataset, we chose to remove the $<2\%$ of edits that fell into this category. 

\paragraph{CodeBERT Threshold.} We perform a final filtering round of edits by validating that each edit snippet identified is relevant to the original outcome by using a CodeBERT score threshold of 0.68. We also allow for a score of 0 to capture entire removals of edits. Preliminary experiments showed that this threshold was achieved by 67.68\% of all initially identified extracted code snippets. We then hand-annotated and validated cases on the edge to see if our threshold worked as expected. We also noted that with the cutoff applied, the number of edits extracted reduced, but the number of trajectories remainder similar, indicating an implicit cutoff in trajectories for when snapshots stray too far away from the initial outcome to extract a meaningful edit. 

\subsection{Dataset Quality Validation}
\label{sec:appendix-dataset-validation}
Given the importance of data quality for the code editing task~\cite{github2025evolving}, we performed a validation of the dataset construction pipeline.
We gathered a sample of 125 before-after edit pairs that were the first and last snapshots from a sample of edit trajectories.
These trajectories were sampled to represent important cases: AI completions with large code prefixes or suffixes, in-line completions, multi-line completions, and long trajectories.
We also randomly sampled before-after edit pairs.
These samples were divided evenly across the three languages in the dataset.
We chose to annotate the first and last snapshots from each trajectory to best capture performance drops with drifts over a trajectory. 

Each trajectory's code edits was manually annotated by an author and compared to the pipeline's. We found that the approach detected the presence of 94.7\% of code edits, with an 88.1\% overlap with code-based edits manually extracted by humans.

\subsection{Dataset Statistics}
\label{sec:appendix-dataset-statistics}
We discuss \dataset in more detail:

\paragraph{Development Context.}
The before-after code pairs were written in Python ($40,390$), JavaScript ($7,993$), and TypeScript ($5,231$).
The edit trajectories also spanned a diverse set of software development task contexts, such as data analysis and visualization ($1,285$), web development ($1,179$), machine learning ($901$), web scraping ($706$), user interfaces ($609$), file processing ($451$), backend development ($371$), game development ($204$), and testing ($125$).

\paragraph{Edit Trajectories.}
The edit trajectories in \dataset lasted for a median duration of 49.7 minutes. 
The minimum edit duration is one second, while the maximum duration was 283.5 days.
In addition, the number of snapshots per trajectory ranged from one to 424 snapshots, with a median of 4 snapshots.
Finally, each trajectory contained a substantial amount of code edits between the initial AI completion and final state (median of 329 Levenshtein distance).

\paragraph{AI Completions.}
Most AI completions were substantial in size, with a median length of 97 characters or 9 lines of code. 
In addition, the code context surrounding the AI completion had a median of 2,949 characters.
The AI completions were invoked by developers via both natural language comments ($697$) and code ($5,134$).

\section{How Do Developers Edit AI-Generated Code?}

\begin{figure*}[t]
\centering
\small
\begin{align}
\texttt{code\_remaining}(Y^d) &= 1 - \frac{\texttt{removal\_ops}(y_0^d, y_t^d) + \texttt{replacement\_ops}(y_0^d, y_t^d)}{\texttt{len}(y_0^d)} \label{eq:code-remaining} \\[4pt]
\texttt{code\_added}(Y^d) &= \frac{\texttt{addition\_ops}(y_0^d, y_t^d)}{\texttt{len}(y_t^d)} \label{eq:code-added} \\[4pt]
\texttt{edit\_volume}(t) &= \sum_{d} \texttt{levenshtein}(y_{t-1}^d, y_{t}^d) \label{eq:edit-volume} \\[4pt]
\texttt{edit\_position}(y_t^d) &= \frac{\texttt{op\_index}(y_{t-1}^d, y_t^d)}{\texttt{len}(y_{t-1}^d)} \label{eq:edit-position}
\end{align}
\caption{\textbf{Metrics for characterizing developer edits to AI-generated code.} All metrics are computed from Levenshtein operations between snapshots of an edit trajectory $Y^d$, where $y_0^d$ is the original AI completion and $y_t^d$ is the code at step $t$.}
\label{fig:metrics}
\end{figure*}

\subsection{Metrics}
\label{sec:appendix-developer-metrics}

\paragraph{Amount of AI-Generated Code Remaining.}
This metric represents the proportion of an AI completion that is remaining at the end of an edit trajectory.
We use the number of Levenshtein removal and replacement operations to identify the original code that was removed or replaced by the developer (\cref{eq:code-remaining}, \cref{fig:metrics}).

\paragraph{Amount of Developer-Added Code.}
This metric represents the proportion of the code at the end of the edit sequence that was newly added by the developer and was not a part of the original AI completion.
We use the Levenshtein addition operations to identify which code was newly added by the developer (\cref{eq:code-added}).

\paragraph{Total Amount of Code Edits.}
This metric represents the total amount of edits made by all developers at a specific time step.
It is computed by aggregating Levenshtein distances made within the same time period (\cref{eq:edit-volume}).

\paragraph{Code Edit Position.}
This metric represents where within a code snippet an edit occurs.
We use the index of a Levenshtein operation to identify the edit's location (\cref{eq:edit-position}).

\begin{figure}[htbp]
    \centering
   \includegraphics[trim=50 650 1300 0, clip, width=\linewidth, keepaspectratio, page=6]{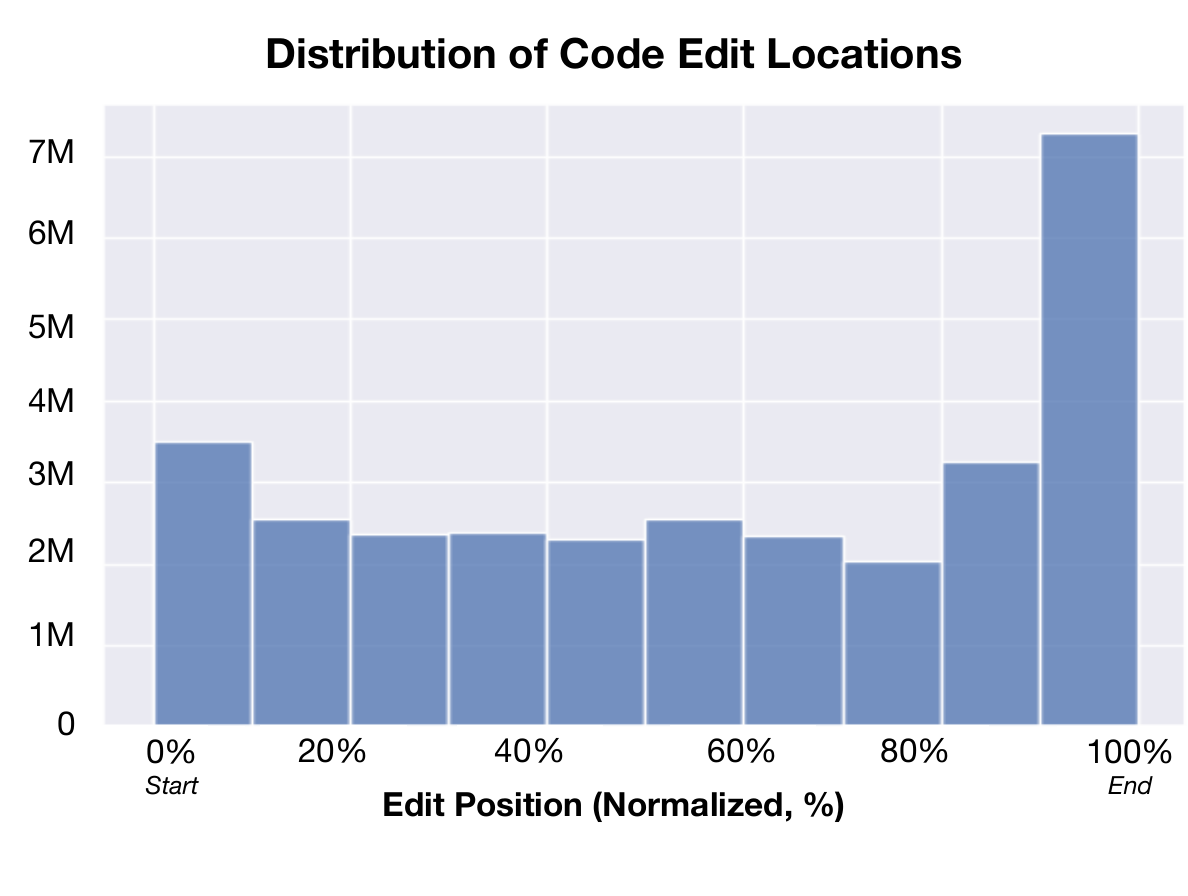} 
    \caption{The distribution of where code edits occur in AI completions.}
    \label{fig:where-edits}
\end{figure}

\subsection{Additional Analyses}

\paragraph{Developers edit most commonly edit at the end of completions.}
The locations in which code edits are performed are approximately uniformly distributed, with increases in frequency at the beginning and end of AI code completions.
This suggests that a significant portion of the code edits are related to integrating the completion into existing code or extending the completion's functionality.
Where edits are made in the code also do not vary over time. 
We did not observe a strong relationship between code edit positions and time ($r=-0.12$, $p<0.0001$) based on Pearson's $r$.

\paragraph{Developers follow a common process to edit AI-generated code across models.}
Different types of edits occurred at varying times after acceptance with statistical significance based on a Kruskal-Wallis test ($H=394.5$, $p < 0.001$)~\cite{mckight2010kruskal}.
Edits related to removing the completion occurred first ($\mu=23.6$ minutes).
This was then followed by edits related to code quality ($\mu=28.3$ minutes), customizing the code ($\mu=49.3$ minutes), and modifying functionality ($\mu=59.2$ minutes).
This indicates that after accepting an AI completion, developers first decide whether they want to keep the code.
After deciding to do so, they then fix any errors in the code, adapt the code to their needs, and finally modify the code's functionality.

\subsection{Types of Code Edits}
\label{sec:appendix-developer-code-edits}
To identify types of code edits, the authors manually annotated every single code edit in 30 trajectories (\ieCustom 85 total snapshots) to get the initial types of edits.
Through discussion between authors, the types and definitions of edits were refined.
For examples of code edits in \dataset, refer to~\cref{tab:edit-types}.

To scale the analysis, we used LLM-as-a-judge to classify edit types (see~\cref{sec:prompt-label-edit-type}).
We labeled each edit snapshot using \modelName{gpt-5-mini} with a 94\% agreement rate with human raters, based on a sample of 120 before-after pairs.

\begin{table*}[h]
    \centering
    \small
    \begin{tabular}{p{2cm}|p{5.5cm}|p{5.5cm}}
    \toprule
        \textbf{Edit Type} & \textbf{Accepted Completion}  & \textbf{New Edit} \\
        \midrule
        Changing Code Functionality &  \begin{lstlisting}
kkk++;
for(let i=0; i<k; i++)\{
        \end{lstlisting} & \begin{lstlisting}
k+=2;
if(k==6)\{
    continue;
        \end{lstlisting}\\
        \midrule
        Improving Code Quality &  \begin{lstlisting}
return quick_sort(left) + 
    middle + 
    quick_sort(right
\end{lstlisting} & \begin{lstlisting}
return quick_sort(left) + 
    middle + 
    quick_sort(right)
        \end{lstlisting}\\
        \midrule
        Customizing Code & \begin{lstlisting}
json.dump(output, f, 
    indent=4, default=str)
\end{lstlisting} & \begin{lstlisting}
json.dump(output_list, f, 
    indent=4, default=str)
        \end{lstlisting}\\
        \midrule
        Deleting Code & \begin{lstlisting}
ax.set_xlabel(
    'No. of Connections'
)
ax.set_ylabel(
    'Network Protocols'
)
ax.set_title('Occurrences')
ax.legend()
\end{lstlisting} & \begin{lstlisting}
<CODE DELETED>
        \end{lstlisting} \\
    \bottomrule
    \end{tabular}
    \caption{Examples of each edit type. We show the original accepted completion and the edit to the completion. For brevity, we shorten the completion and edits.}
    \label{tab:edit-types}
\end{table*}

\section{How Well Do LLMs Learn from Developer Edits?}

\subsection{Experimental Setup}
\label{sec:appendix-modeling-setup}

\paragraph{Task.}
In addition to the generation task described in \S\ref{sec:experiments}, we also trained a model to predict whether the AI completion $y_0$ would be deleted, unmodified, or modified, based the amount of AI-generated code remaining (see \S\ref{sec:data-analysis-methodology}).

\paragraph{Datasets.}
Trajectories where the final code snippet was empty were removed from consideration for the generation task.
This resulted in a slightly smaller dataset of 4,710 trajectories. 
We opted to model trajectories that did not result in complete deletions because it is a harder task to train models to not generate any new code at all.
For all tasks, the train, development, and tests sets were created corresponding to 80\%, 10\%, and 10\% of developers respectively (randomly split).

For the generation task, we performed additional processing of the trajectories. 
To ensure that the trajectories represented the same code editing task, we further truncated each edit trajectory to the snapshots that occurred within the first 10 minutes, as this captured a majority of edit edit volume (\S\ref{sec:data-analysis-results})).
Each edit trajectory was split into before-after pairs, where the before state was the original code completion or intermediate code edit snapshot and the end state was the final code edit snapshot in the trajectory; this resulted in a dataset of $23,874$ before-after pairs.

\smallDataset was created by a study author who had extensive Python programming experience by selecting 15 problems from HumanEval~\citep{chen2021evaluating} and manually perturbing them to fit an edit described by modifying functionality, customizing code, or improving code quality.
The 15 problems were distributed equally across all edit types.
The quality of the editing problems were reviewed and approved by an independent experienced Python programmer.

\paragraph{Training and Inference Hyper-Parameters.}
We assessed the effectiveness of both parameter-efficient fine-tuning (LoRA) using \dataset by fine-tuning \modelName{Qwen2.5-Coder} at 3B and 7B parameters, \modelName{Qwen3} at 4B and 8B parameters, \modelName{Llama3.2} at 3B parameters, and \modelName{Gemma-3} at 4B parameters.
We used LoRA fine-tuning as it showed better performance in preliminary experiments.
Tasks were learned with a causal language modeling head.
The models were trained for 3 epochs for the generation task on NVIDIA A 100-80GB and L40S GPUs using the AdamW optimizer~\cite{kingma2015adam}, a batch size of 32, warm-up of 0.03, LoRA alpha of 32, LoRA rank of 16, BF16 precision, and maximum context lengths of 4096 for the generation task. 
We selected the best model based on a learning rate sweep covering $[1e-5, 5e-5, 5e-6]$ identified different learning rates as optimal based on the base model family as well as the task at hand.
Consequently, we selected those learning rates for all subsequent experiments.

For the fine-tuning experiments for the generation task, the models had the following hyperparameters:
\begin{itemize}
    \item Fine-tuned \modelName{Qwen2.5-Coder-3B}: $5e-5$
    \item Fine-tuned \modelName{Qwen2.5-Coder-7B}: $1e-5$
    \item Fine-tuned \modelName{Qwen3-4B}: $5e-5$
    \item Fine-tuned \modelName{Qwen3-8B}: $5e-5$
    \item Fine-tuned \modelName{Llama-3.2-3B}: $1e-5$
    \item Fine-tuned \modelName{Gemma3-4B-IT}: $1e-5$
\end{itemize}

\paragraph{Decoding Strategy.}
Code generation is known to benefit from sampling~\cite{li2022competition, shypula2024learning}, so inference in the generation task specifically was performed with beam sampling with 3 beams, a temperature of 0.7, default top-k and top-p values, and early stopping. 

\paragraph{Model Access.} For the large base models, we used OpenRouter for access to all other models.
For every model provider, the default settings were used.

\paragraph{Evaluation Metrics.}
For the generation task, text similarity metrics (Hugging Face Version 0.4.0) and normalized Levenshtein similarity were reported. The F1 score and Levenshtein similarity were used to determine the best checkpoint for evaluation.

\paragraph{LLM Judge for Evaluating Edits.} 
We use \modelName{Claude-Sonnet-5} as an LLM judge to evaluate the generations for \smallDataset code editing problems.
The judge is validated by randomly selecting 25 model generations, and having a human annotator separately annotate each generation using the same rubrics given the model.
This procedure yields Krippendorff's $\alpha = 0.84$ between an LLM judge and one human rater, indicating reliable agreement~\cite{krippendorff1989content}.

\section{Prompts}
Here we report the prompts utilized for various tasks during the dataset analysis, labeling, and model training stages. 

\subsection{Prompt to Identify Code Context}
We replicated the multi-step prompting process employed by Copilot Arena ~\cite{chi2025copilot} to cluster code contexts and categorize code trajectories. Below are the three prompts used directly from Copilot Arena with model \modelName{GPT-4o-mini}.
Refer to~\cref{fig:code-context-prompt-1}, ~\cref{fig:code-context-prompt-2}, and ~\cref{fig:code-context-prompt-3} for the prompts.

\begin{figure*}
    \centering
    \begin{tcolorbox}[colback=white,colframe=black!70,boxrule=0.8pt,arc=3mm]
\ttfamily
\textbf{System Prompt}

You are a helpful assistant that describes code files in a single,
concise sentence. Focus on the main purpose and functionality of the
code. Keep descriptions clear, technical, and under 100 characters. Do
not mention file names or extensions in your description.

\vspace{0.5em}
\textbf{General Prompt}
Describe this code in one sentence
\end{tcolorbox}
    
    \caption{The first prompt used to identify the code context, which summarizes the code file.}
    \label{fig:code-context-prompt-1}
\end{figure*}

\begin{figure*}
    \centering
\begin{tcolorbox}[colback=white,colframe=black!70,boxrule=0.8pt,arc=3mm]
\ttfamily
\textbf{General Prompt}

You are a code organization expert.

Analyze the provided code descriptions and:
\begin{enumerate}
\item Identify 5-10 main functional clusters or themes
\item Assign each description to the most appropriate cluster
\item Provide a brief name and description for each cluster
\item Format the response as valid JSON with the following structure:
\end{enumerate}

\begin{verbatim}
    {
      "clusters": [
        {
          "name": "cluster_name",
          "description": "brief cluster description",
          "descriptions": ["description", "description2"]
        }
      ]
    }
\end{verbatim}
\end{tcolorbox}
    
    \caption{The second prompt used to identify the code context, which generates clusters of code files.}
    \label{fig:code-context-prompt-2}
\end{figure*}

\begin{figure*}
    \centering
\begin{tcolorbox}[colback=white,colframe=black!70,boxrule=0.8pt,arc=3mm]
\ttfamily
\textbf{System Prompt}

Please categorize the following code into one of these categories:

\begin{itemize}
\item User Interaction and Input Handling: Code that manages user inputs, prompts, and basic interaction with the system
\item Frontend Development and UI Design: Code snippets focused on designing user interfaces and creating interactive components
\item Backend Development and APIs: Server-side logic, data management, and API integration for applications
\item Algorithm Design and Problem Solving: Code implementing algorithms to solve computational problems or optimize tasks
\item Data Processing and File Operations: Code that reads, writes, or processes data from files and other data sources
\item Game Development and Simulations: Code focused on creating games, simulations, and managing game dynamics
\item Artificial Intelligence and Machine Learning: Code related to AI and machine learning for training, inference, and application
\end{itemize}

\vspace{0.5em}
\textbf{General Prompt}

Only respond with the exact category name that best fits. No other text.

Here’s the code:

[code content]
\end{tcolorbox}
    
    \caption{The third prompt used to identify the code context, which classifies each code file into categories.}
    \label{fig:code-context-prompt-3}
\end{figure*}

\subsection{Prompt to Label Edit Type}
\label{sec:prompt-label-edit-type}
In order to categorize edits into one of four categories (changing code functionality, improving code quality, customizing code, deleting code), we used \modelName{GPT-5-mini} for labeling with the following prompt. 
Refer to~\cref{fig:edit-type-prompt-1}, ~\cref{fig:edit-type-prompt-2}, and ~\cref{fig:edit-type-prompt-3} for the prompts.

\begin{figure*}
    \begin{tcolorbox}[colback=white,colframe=black!70,boxrule=0.8pt,arc=3mm]
\ttfamily
\textbf{General Prompt}

You are a code-edit classifier. You must determine how each line in current\_snippet changed relative to earlier versions of the code.

\vspace{0.5em}
\textbf{Inputs}

You will receive:
\begin{itemize}
\item outcome\_snippet: the oldest known version of the snippet
\item prev\_full\_snippet: the version immediately before the current edit. it may or may not simply be the outcome\_snippet itself
\item current\_full\_snippet: the complete snippet at the current moment
\item current\_snippet: ONLY the lines that changed in this edit (a subset of current\_full\_snippet)
\item after\_full\_snippet: the next version after the current edit.  it may or may not simply be the current\_full\_snippet itself
\end{itemize}

\vspace{0.5em}
\textbf{Timeline of code from oldest to newest:}

outcome\_snippet, prev\_full\_snippet, current\_full\_snippet, after\_full\_snippet

\vspace{0.5em}
\textbf{Your job}

For each line in current\_snippet, assign exactly one label using the taxonomy below.

Prioritize using prev\_full\_snippet to understand the change. Use all surrounding snippets only to understand how the line changed. Prioritize using prev\_full\_snippet to understand how the line changed - only when you cannot do this use the other snippets as well.

\vspace{0.5em}
\textbf{Taxonomy}

\begin{enumerate}
\item \textbf{customizing-personalizing}: The line keeps the same behavior but changes literal values or symbol names that customizes the code to fit the remaining file well, stylistically or otherwise. Examples: renaming variables, changing constant values/literal values, swapping variable names, replacing variables with literals, changing variable values. Changing the value of a comment string is improving-code-quality instead.
\item \textbf{improving-code-quality}: The line improves readability, formatting, comments, or syntax without changing semantic meaning.
Examples: Includes fixing syntax such as closing code structures, adjusting tabbing, cleaning up the code (such as removing unnecessary comments, adding new empty lines), deleting partial code, and adding comments. If the entire line was removed, use start-over-commenting-out instead.
\end{enumerate}
\end{tcolorbox}
\caption{The prompt used to edit label types (part 1). The remainder of the prompt is located in \cref{fig:edit-type-prompt-2} and \cref{fig:edit-type-prompt-3}.}
    \label{fig:edit-type-prompt-1}
\end{figure*}

\begin{figure*}
    \begin{tcolorbox}[colback=white,colframe=black!70,boxrule=0.8pt,arc=3mm]
\ttfamily
\begin{enumerate}
\item \textbf{modifying-functionality}: The line introduces new code that changes program behavior. Examples: Includes adding methods, changing code logic (such as adding new method calls, changing control flow), and changing API calls (such as changing the API, API methods used, parameters passed in). Do not use if only names/values were changed but the logic remains the same, that is customizing-personalizing instead.
\item \textbf{start-over-commenting-out}: The line corresponds to complete deletion of an earlier line OR the line in current\_snippet is an empty string representing removed code. This includes lines that were fully commented out.
Partial deletions that still leave a meaningful line are improving-code-quality instead.
If a line is empty but not due to code deletion, rather the empty line is a new addition, then this is improving-code-quality instead. If the entire snippet of code in current\_snippet is an empty string or not provided, label the empty string under this category. If a new empty line is added but has a matching empty line in prev\_full\_snippet, then this is unchanged instead.
\item \textbf{unchanged}: The line is EXACTLY the same between prev\_full\_snippet and current\_full\_snippet. Even one character difference means it is NOT unchanged.
\end{enumerate}

\textbf{Tie-breaking:}

If a line seems ambiguous, apply this priority order: modifying-functionality $>$ customizing-personalizing $>$ improving-code-quality $>$ start-over-commenting-out $>$ unchanged

\vspace{0.5em}
\textbf{Output Format}

Return a JSON array, where each object includes:
\begin{verbatim}
{
    "line": "<the line from current_snippet>",
    "label": "<one label from the taxonomy>"
}
\end{verbatim}

Use strict JSON (no comments, no trailing commas, no code fences).

Here are some examples to guide you. I have provided additional rationale per label, but you don't need to include that in your output.

\vspace{0.5em}
\textbf{Example 1:}

outcome\_snippet: \{outcome\_snippet\_1\}

current\_full\_snippet: \{current\_full\_snippet\_1\}

current\_snippet: \{current\_snippet\_1\}

prev\_full\_snippet: \{prev\_full\_snippet\_1\}

after\_full\_snippet: \{after\_full\_snippet\_1\}

output: \{output\_1\}

\vspace{0.5em}
\end{tcolorbox}

\caption{The prompt used to edit label types (part 2). The remainder of the prompt is located in \cref{fig:edit-type-prompt-1} and \cref{fig:edit-type-prompt-3}.}
\label{fig:edit-type-prompt-2}
\end{figure*}

\begin{figure*}
    \begin{tcolorbox}[colback=white,colframe=black!70,boxrule=0.8pt,arc=3mm]
\ttfamily

\textbf{Example 2:}

outcome\_snippet: \{outcome\_snippet\_2\}

current\_full\_snippet: \{current\_full\_snippet\_2\}

current\_snippet: \{current\_snippet\_2\}

prev\_full\_snippet: \{prev\_full\_snippet\_2\}

after\_full\_snippet: \{after\_full\_snippet\_2\}

output: \{output\_2\}

\vspace{0.5em}

\textbf{Example 3:}

outcome\_snippet: \{outcome\_snippet\_3\}

current\_full\_snippet: \{current\_full\_snippet\_3\}

current\_snippet: \{current\_snippet\_3\}

prev\_full\_snippet: \{prev\_full\_snippet\_3\}

after\_full\_snippet: \{after\_full\_snippet\_3\}

output: \{output\_3\}

\vspace{0.5em}

\textbf{Example 4:}

outcome\_snippet: \{outcome\_snippet\_4\}

current\_full\_snippet: \{current\_full\_snippet\_4\}

current\_snippet: \{current\_snippet\_4\}

prev\_full\_snippet: \{prev\_full\_snippet\_4\}

after\_full\_snippet: \{after\_full\_snippet\_4\}

output: \{output\_4\}

\vspace{0.5em}

\textbf{Example 5:}

outcome\_snippet: \{outcome\_snippet\_5\}

current\_full\_snippet: \{current\_full\_snippet\_5\}

current\_snippet: \{current\_snippet\_5\}

prev\_full\_snippet: \{prev\_full\_snippet\_5\}

after\_full\_snippet: \{after\_full\_snippet\_5\}

output: \{output\_5\}
    \end{tcolorbox}
    \caption{The prompt used to edit label types (part 3). The remainder of the prompt is located in \cref{fig:edit-type-prompt-1} and \cref{fig:edit-type-prompt-2}.}
    \label{fig:edit-type-prompt-3}
\end{figure*}

\subsection{Prompt for Generation Task.}
For the prompt provided during fine-tuning and inference for the generation task, refer to \cref{fig:generation-task-prompt}.

\begin{figure*}
    \begin{tcolorbox}[
    colback=white,
    colframe=black!70,
    boxrule=0.8pt,
    arc=3mm,
]
\ttfamily
\textbf{General Prompt}

You are a code-edit prediction model.

\vspace{0.5em}
\textbf{Given} 

COMPLETION (snippet of code inserted by the user), PREFIX (code before the COMPLETION), and SUFFIX (code after the COMPLETION).

\vspace{0.5em}
\textbf{Task}

Predict what the COMPLETION will look like after the user makes their final edits later.

\vspace{0.5em}
\textbf{Rules}
\begin{itemize}
\item Output ONLY the final edited COMPLETION (no prefix/suffix, no explanations, no markdown fences)
\item Preserve semantics; make the minimal necessary edits
\item Keep the same language and style as the surrounding code
\item If the completion would be completely deleted in the final version, output exactly: \texttt{<EDIT DELETED>}
\end{itemize}

\vspace{0.5em}
\textbf{Example 1}

PREFIX: \{prefix\_1\}

COMPLETION: \{completion\_1\}

SUFFIX: \{suffix\_1\}

FINAL EDITED VERSION OF COMPLETION SNIPPET: \{target\_0\}

\vspace{0.5em}
\textbf{Example 2}

PREFIX: \{prefix\_2\}

COMPLETION: \{completion\_2\}

SUFFIX: \{suffix\_2\}

FINAL EDITED VERSION OF COMPLETION SNIPPET: \{target\_1\}

\vspace{0.5em}
\textbf{Your Turn}

PREFIX: \{prefix\}

COMPLETION: \{completion\}

SUFFIX: \{suffix\}

\vspace{0.5em}
Here are some initial edits to the COMPLETION (intermediate versions, oldest to newest).  
Use them as additional context to predict the FINAL edited completion.

If an intermediate edit is \texttt{<EDIT DELETED>}, it does not mean the final output is \texttt{<EDIT DELETED>}.

\vspace{0.5em}

EDIT 1: \{edit\_1\}

EDIT 2: \{edit\_2\}

\vspace{0.5em}
\textbf{Final Edited Version of Completion Snippet:}

\end{tcolorbox}
\caption{The prompt provided during fine-tuning and inference for the generation task.}
\label{fig:generation-task-prompt}
\end{figure*}

\subsection{Prompt to Judge LLM Generations}
For the prompt provided to judge LLM generations for code edits from \smallDataset, refer to Figure~\ref{fig:judge-prompt}.

\begin{figure*}
    \begin{tcolorbox}[
    colback=white,
    colframe=black!70,
    boxrule=0.8pt,
    arc=3mm,
]
\ttfamily
\textbf{General Prompt}

You are a code-edit prediction model.

\vspace{0.5em}
\textbf{Given} 

COMPLETION (snippet of code inserted by the user), PREFIX (code before the COMPLETION), and SUFFIX (code after the COMPLETION).

\vspace{0.5em}
\textbf{Task}

Predict what the COMPLETION will look like after the user makes their final edits later.

\vspace{0.5em}
\textbf{Rules}
\begin{itemize}
\item Output ONLY the final edited COMPLETION (no prefix/suffix, no explanations, no markdown fences)
\item Preserve semantics; make the minimal necessary edits
\item Keep the same language and style as the surrounding code
\item If the completion would be completely deleted in the final version, output exactly: \texttt{<EDIT DELETED>}
\end{itemize}

\vspace{0.5em}
\textbf{Example 1}

PREFIX: \{prefix\_1\}

COMPLETION: \{completion\_1\}

SUFFIX: \{suffix\_1\}

FINAL EDITED VERSION OF COMPLETION SNIPPET: \{target\_0\}

\vspace{0.5em}
\textbf{Example 2}

PREFIX: \{prefix\_2\}

COMPLETION: \{completion\_2\}

SUFFIX: \{suffix\_2\}

FINAL EDITED VERSION OF COMPLETION SNIPPET: \{target\_1\}

\vspace{0.5em}
\textbf{Your Turn}

PREFIX: \{prefix\}

COMPLETION: \{completion\}

SUFFIX: \{suffix\}

\vspace{0.5em}
Here are some initial edits to the COMPLETION (intermediate versions, oldest to newest).  
Use them as additional context to predict the FINAL edited completion.

If an intermediate edit is \texttt{<EDIT DELETED>}, it does not mean the final output is \texttt{<EDIT DELETED>}.

\vspace{0.5em}

EDIT 1: \{edit\_1\}

EDIT 2: \{edit\_2\}

\vspace{0.5em}
\textbf{Final Edited Version of Completion Snippet:}

\end{tcolorbox}
\caption{The prompt provided during fine-tuning and inference for the generation task.}
\label{fig:generation-task-prompt}
\end{figure*}

\begin{figure*}
    \begin{tcolorbox}[
    colback=white,
    colframe=black!70,
    boxrule=0.8pt,
    arc=3mm,
]
\ttfamily
\textbf{General Prompt}

You are a strict grading assistant for a code-edit prediction task.

\vspace{0.5em}
A model was given COMPLETION (a snippet of code inserted by the user), PREFIX (code before the COMPLETION), and SUFFIX (code after the COMPLETION), and asked to predict what the COMPLETION would look like after the user's final edits. You will grade its FINAL GENERATION against a rubric.

\vspace{0.5em}

PREFIX: \{prefix\}

COMPLETION: \{completion\}

SUFFIX: \{suffix\}

\vspace{0.5em}
Here are some initial edits to the COMPLETION (intermediate versions, oldest to newest).  
Use them as additional context to predict the final label.

If an intermediate edit is \texttt{<EDIT DELETED>}, it does not necessarily mean the final completion is \texttt{<EDIT DELETED>}.

EDIT 1: \{edit\_1\}

\vspace{0.5em}
FINAL GENERATION (the model's predicted final edited COMPLETION): \{llm\_generation\}

DIFF (COMPLETION -> FINAL GENERATION):
\{code\_diff\}

\vspace{0.5em}
Grade the FINAL GENERATION against each rubric item below. For each item, choose exactly one of its listed score levels.

Item \{rubric\_id\}: \{rubric\_criteria\}

\vspace{0.5em}
Respond with ONLY a JSON array, one object per item above in order, each with keys "item" (the item number), "score" (the integer score you chose, matching one of its listed score levels), and "justification" (a one-sentence justification). No other text, no markdown fences.

\vspace{0.5em}
\end{tcolorbox}
\caption{The prompt provided for judging model generations on \smallDataset.}
\label{fig:judge-prompt}
\end{figure*}

\end{document}